\pdfoutput=1

\documentclass[twocolumn,showpacs,aps,prl,superscriptaddress]{revtex4}

\usepackage{graphicx}
\usepackage{dcolumn}
\usepackage{amsmath}
\usepackage{epsfig}
\usepackage{hhline}
\usepackage{soul}
\usepackage{tabularx,pifont,multirow}
\usepackage{colordvi}
\usepackage{bm} % for bold letters \bm{\beta}
\usepackage{amssymb}
\usepackage[usenames,dvipsnames]{xcolor}
\definecolor{taplum}{rgb}{0.67843, 0.49804, 0.65882}
\RequirePackage{xspace}

\usepackage{xspace} 
\usepackage{upgreek}

\def\lhcb {\mbox{LHCb}\xspace}

\def\lhc    {\mbox{LHC}\xspace}

\def\MagUp {\mbox{\em Mag\kern -0.05em Up}\xspace}

 \def\PLambda     {\ensuremath{\Lambda}\xspace}                 
                  
 \def\Pnu         {\ensuremath{\nu}\xspace}                 
                  
 \def\Ppi         {\ensuremath{\pi}\xspace}

 \def\Ptau        {\ensuremath{\tau}\xspace}                 
 \def\Pds         {\ensuremath{\D_\squark}\xspace}

 \def\PB      {\ensuremath{B}\xspace}                 
                  
 \def\PD      {\ensuremath{D}\xspace}

 \def\PK      {\ensuremath{K}\xspace}

 \def\Pc      {\ensuremath{c}\xspace}                 
                  
 \def\Pe      {\ensuremath{e}\xspace}

 \def\Pi      {\ensuremath{i}\xspace}

 \def\Pp      {\ensuremath{p}\xspace}

 \def\Ps      {\ensuremath{s}\xspace}

\def\epem       {{\ensuremath{\Pe^+\Pe^-}}\xspace}

\def\taup       {{\ensuremath{\Ptau^+}}\xspace}
\def\taum       {{\ensuremath{\Ptau^-}}\xspace}

\def\neu        {{\ensuremath{\Pnu}}\xspace}
\def\neub       {{\ensuremath{\overline{\Pnu}}}\xspace}

\def\neut       {{\ensuremath{\neu_\tau}}\xspace}
\def\neutb      {{\ensuremath{\neub_\tau}}\xspace}

\def\W      {{\ensuremath{\PW}}\xspace}

\def\squark    {{\ensuremath{\Ps}}\xspace}

\def\cquark    {{\ensuremath{\Pc}}\xspace}

\def\pion   {{\ensuremath{\Ppi}}\xspace}

\def\pip    {{\ensuremath{\pion^+}}\xspace}
\def\pim    {{\ensuremath{\pion^-}}\xspace}

  \def\Kbar    {{\kern 0.2em\overline{\kern -0.2em \PK}{}}\xspace}

\def\KorKbar    {\kern 0.18em\optbar{\kern -0.18em K}{}\xspace}

  \def\Dbar    {{\kern 0.2em\overline{\kern -0.2em \PD}{}}\xspace}
\def\D       {{\ensuremath{\PD}}\xspace}

\def\DorDbar    {\kern 0.18em\optbar{\kern -0.18em D}{}\xspace}

\def\Dp      {{\ensuremath{\D^+}}\xspace}

\def\Ds      {{\ensuremath{\D^+_\squark}}\xspace}
\def\Dsp     {{\ensuremath{\D^+_\squark}}\xspace}

\def\B       {{\ensuremath{\PB}}\xspace}
\def\Bbar    {{\ensuremath{\kern 0.18em\overline{\kern -0.18em \PB}{}}}\xspace}

\def\BorBbar    {\kern 0.18em\optbar{\kern -0.18em B}{}\xspace}

\def\Bu      {{\ensuremath{\B^+}}\xspace}

\def\Bp      {{\ensuremath{\Bu}}\xspace}

  \def\Y#1S{\ensuremath{\PUpsilon{(#1S)}}\xspace}% no space before {...}!

\def\proton      {{\ensuremath{\Pp}}\xspace}

\def\Lz          {{\ensuremath{\PLambda}}\xspace}
\def\Lbar        {{\ensuremath{\kern 0.1em\overline{\kern -0.1em\PLambda}}}\xspace}
\def\LorLbar    {\kern 0.18em\optbar{\kern -0.18em \PLambda}{}\xspace}

\def\Lcp      {{\ensuremath{\Lz^+_\cquark}}\xspace}

\def\to                 {\ensuremath{\rightarrow}\xspace}

\def\AT#1     {\ensuremath{A_{\mathrm{T}}^{#1}}\xspace}           % 2

\def\C#1      {\ensuremath{\mathcal{C}_{#1}}\xspace}                       % 9
\def\Cp#1     {\ensuremath{\mathcal{C}_{#1}^{'}}\xspace}                    % 7
\def\Ceff#1   {\ensuremath{\mathcal{C}_{#1}^{\mathrm{(eff)}}}\xspace}        % 9  
\def\Cpeff#1  {\ensuremath{\mathcal{C}_{#1}^{'\mathrm{(eff)}}}\xspace}       % 7
\def\Ope#1    {\ensuremath{\mathcal{O}_{#1}}\xspace}                       % 2
\def\Opep#1   {\ensuremath{\mathcal{O}_{#1}^{'}}\xspace}                    % 7

\newcommand{\tev}{\ensuremath{\mathrm{\,Te\kern -0.1em V}}\xspace}
\newcommand{\gev}{\ensuremath{\mathrm{\,Ge\kern -0.1em V}}\xspace}
\newcommand{\mev}{\ensuremath{\mathrm{\,Me\kern -0.1em V}}\xspace}
\newcommand{\kev}{\ensuremath{\mathrm{\,ke\kern -0.1em V}}\xspace}
\newcommand{\ev}{\ensuremath{\mathrm{\,e\kern -0.1em V}}\xspace}
\newcommand{\gevc}{\ensuremath{{\mathrm{\,Ge\kern -0.1em V\!/}c}}\xspace}
\newcommand{\mevc}{\ensuremath{{\mathrm{\,Me\kern -0.1em V\!/}c}}\xspace}
\newcommand{\gevcc}{\ensuremath{{\mathrm{\,Ge\kern -0.1em V\!/}c^2}}\xspace}
\newcommand{\gevgevcccc}{\ensuremath{{\mathrm{\,Ge\kern -0.1em V^2\!/}c^4}}\xspace}
\newcommand{\mevcc}{\ensuremath{{\mathrm{\,Me\kern -0.1em V\!/}c^2}}\xspace}

\def\cm   {\ensuremath{\mathrm{ \,cm}}\xspace}

\def\mm   {\ensuremath{\mathrm{ \,mm}}\xspace}

\def\mum  {\ensuremath{{\,\upmu\mathrm{m}}}\xspace}

\def\mub{\ensuremath{{\mathrm{ \,\upmu b}}}\xspace}

\def\sec  {\ensuremath{\mathrm{{\,s}}}\xspace}

\def\gsim{{~\raise.15em\hbox{$>$}\kern-.85em
          \lower.35em\hbox{$\sim$}~}\xspace}
\def\lsim{{~\raise.15em\hbox{$<$}\kern-.85em
          \lower.35em\hbox{$\sim$}~}\xspace}

\def\mrad{\ensuremath{\mathrm{ \,mrad}}\xspace}
\def\rad{\ensuremath{\mathrm{ \,rad}}\xspace}

\def\evtgen     {\mbox{\textsc{EvtGen}}\xspace}

\def\pythia     {\mbox{\textsc{Pythia}}\xspace}

\def\tell1  {TELL1\xspace}
\def\ukl1   {UKL1\xspace}

\newcommand{\eg}{\mbox{\itshape e.g.}\xspace}

\def\W               {{\ensuremath{\rm W}}\xspace}
\def\Si               {{\ensuremath{\rm Si}}\xspace}
\def\Ge              {{\ensuremath{\rm Ge}}\xspace}
\newcommand{\pot}{\ensuremath{\mathrm{\,PoT}}\xspace}

\def\murad  {\ensuremath{{\,\upmu\mathrm{rad}}}\xspace}

\newcommand{\tevc}{\ensuremath{{\mathrm{\,Te\kern -0.1em V\!/}c}}\xspace}
\newcommand{\tevtevcccc}{\ensuremath{{\mathrm{\,Te\kern -0.1em V^2\!/}c^4}}\xspace}
\newcommand{\gevgevcc}{\ensuremath{{\mathrm{\,Ge\kern -0.1em V^2\!/}c^2}}\xspace}
\newcommand{\tevtevcc}{\ensuremath{{\mathrm{\,Te\kern -0.1em V^2\!/}c^2}}\xspace}

\def\xaxis   {{\ensuremath{X}}\xspace}
\def\yaxis   {{\ensuremath{Y}}\xspace}
\def\zaxis   {{\ensuremath{Z}}\xspace}
\def\x   {{\ensuremath{\sc x}}\xspace}
\def\y   {{\ensuremath{\sc y}}\xspace}
\def\z   {{\ensuremath{\sc z}}\xspace}

\def\thetay   {{\ensuremath{ \theta_{\y} }}\xspace}
\def\thetayDsTau   {{\ensuremath{ \theta_{\y,\Pds\Ptau} }}\xspace}

\def\thetaxDsTau   {{\ensuremath{ \theta_{\x,\Pds\Ptau} }}\xspace}

\def\at   {{\ensuremath{ a }}\xspace}
\def\apt   {{\ensuremath{ a' }}\xspace}
\def\dt   {{\ensuremath{ d }}\xspace}
\def\dpt   {{\ensuremath{ d' }}\xspace}
\def\apdt   {{\ensuremath{ a'_d }}\xspace}
\def\deltat   {{\ensuremath{ \delta }}\xspace}

\def\Lc   {{\ensuremath{ L }}\xspace}
\def\thc   {{\ensuremath{ \theta_C }}\xspace}
\def\Ltarc   {{\ensuremath{ L_{\mathrm{tar}} }}\xspace}
\def\phad   {{\ensuremath{ p_{\mathrm 3\pi} }}\xspace}

\def\pythia     {\mbox{\textsc{Pythia}}\xspace}
\def\s0z   {{\ensuremath{ s_{0,z} }}\xspace}
\def\evtgen     {\mbox{\textsc{EvtGen}}\xspace}

\def\taupipipi   {{\ensuremath{ \taup\to\pip\pim\pip\neutb}}\xspace}

\def\Wieta   {{\ensuremath{ {\cal W}_i(\eta) }}\xspace}
\def\Wipmeta   {{\ensuremath{ {\cal W}_i^\pm(\eta) }}\xspace}
\def\Wipeta   {{\ensuremath{ {\cal W}_i^+(\eta) }}\xspace}
\def\Wimeta   {{\ensuremath{ {\cal W}_i^-(\eta) }}\xspace}

\def\Wypeta   {{\ensuremath{ {\cal W}_\yaxis^+(\eta) }}\xspace}
\def\Wymeta   {{\ensuremath{ {\cal W}_\yaxis^-(\eta) }}\xspace}

\long\def\inst#1{\par\nobreak\kern 4pt\nobreak
  {\it #1}\par\vskip 10pt plus 3pt minus 3pt}

\newcommand{\CHa}{}
\newcommand{\CHb}{}
\newcommand{\CHc}{}

\begin{document}

\begin{flushleft}
\end{flushleft}

\title{
{
  \large \bf \boldmath
Novel method for the direct measurement of the \Ptau lepton dipole moments 
}
}

\author{J.~Fu}
\affiliation{INFN Sezione di Milano and Universit\`a di Milano, Milano, Italy}
\author{M.A.~Giorgi}
\affiliation{INFN Sezione di Pisa and Universit\`a di Pisa, Pisa, Italy}
\author{L.~Henry}
\affiliation{IFIC, Universitat de Val\`encia-CSIC, Valencia, Spain}
\author{D.~Marangotto}
\affiliation{INFN Sezione di Milano and Universit\`a di Milano, Milano, Italy}
\author{F.~Mart\'inez Vidal}
\affiliation{IFIC, Universitat de Val\`encia-CSIC, Valencia, Spain}
\author{A.~Merli}
\affiliation{INFN Sezione di Milano and Universit\`a di Milano, Milano, Italy}
\author{N.~Neri}
\affiliation{INFN Sezione di Milano and Universit\`a di Milano, Milano, Italy}
\author{J.~Ruiz Vidal}
\affiliation{IFIC, Universitat de Val\`encia-CSIC, Valencia, Spain}

\date{\today}% It is always \today, today, but you may specify any date with \date.

\begin{abstract}
  A novel method for the direct measurement of the elusive magnetic and
electric dipole moments of the \Ptau lepton is presented.
The experimental approach relies on the production of \taup leptons
from $\Dsp\to\taup\neut$ decays, {\CHa originating} in fixed-target collisions at the \lhc.
A sample of polarized \taup leptons is kinematically selected  
and subsequently channeled in a bent crystal.
The magnetic and electric dipole
moments of the \taup lepton are measured by determining the rotation of the spin-polarization vector
induced by the intense electromagnetic field between crystal atomic planes.
The experimental technique is discussed along with the expected sensitivities.

\end{abstract}
\maketitle

\raggedbottom % tell latex it's ok not to leave blank space in the bottom of the page 

{\CHb Measurements} of the electromagnetic dipole moments for common particles like the electron, muon and nucleons, combined with precise theoretical calculations,
provide stringent tests of physics within and beyond the
Standard Model (SM)~\cite{Mohr:2015ccw,Andreev:2018ayy,Bennett:2006fi,Bennett:2008dy,Afach:2015sja,Schneider:2017,Sahoo:2016zvr,Graner:2016ses}.
For short-lived particles like heavy baryons and the \Ptau lepton,
the
short lifetime
($\sim 10^{-13}\sec$)
prevents
the use of the spin-precession technique adopted in the muon $g-2$ experiment~\cite{Bennett:2006fi,Bennett:2008dy}.
Recently, the possibility of {\CHb directly measuring} the
electromagnetic dipole moments of short-lived baryons, produced in fixed-target collisions
at the Large Hadron Collider (\lhc)
and channeled in bent crystals~\cite{Baryshevsky:2016cul,Botella:2016ksl,Baryshevsky:2017yhk,Bezshyyko:2017var,Bagli:2017foe,Baryshevsky:2018dqq}, has been considered.
For the \Ptau lepton,
the use of $\Bp \to \taup \neut$ decays was suggested~\cite{Samuel:1990su} and more
recently the $\Dsp \to \taup \neut$ process with higher yield has been
explored~\cite{Fomin:2018ybj}.
In this Letter, a novel method that fully exploits the polarization properties
of \taup leptons produced in \Dsp decays is proposed.
The magnetic (MDM) and the electric (EDM) dipole moments are defined
as ${\bm \mu} = g e\hbar/(2m_{\Ptau}c) {\mathbf s}/2$ and ${\bm \delta} = d e\hbar/(2m_{\Ptau}c) {\mathbf s}/2$, respectively, where 
$m_{\Ptau}$ is the
\Ptau mass, $g$ ($d$) is the gyromagnetic (gyroelectric) factor,
and ${\mathbf s}$ is the spin-polarization vector~\cite{Leader2011}.
In the SM, the \Ptau anomalous MDM is expected to be $\at = (g-2)/2 \approx 10^{-3}$~\cite{Eidelman:2007sb}, and its EDM, \dt,
to be minuscule~\cite{Pospelov:2013sca}.
However, the dipole moments can be largely enhanced in {\CHa the} presence of physics beyond the SM~\cite{Pich:2013lsa,Dekens:2018bci}.
Methods based on precise measurements of
{\CHa the}
$\taup\taum$ pair production cross section in $\epem$ annihilations set indirect limits on \at at {\CHb the} few
percent level~\cite{Abdallah:2003xd},
still above the SM prediction, and lead {\CHa to} limits on $\delta$
at $10^{-16}~e\cm$ level~\cite{Inami:2002ah}.
Other indirect measurements have been suggested to improve the precision~\cite{Pich:2013lsa,Hayreter:2015cia,Chen:2018cxt}.

%
%
%
%
% THIS HAS BEEN MOVED TO PART6.
%
%
%

\begin{figure}
\includegraphics[width=0.4\textwidth]{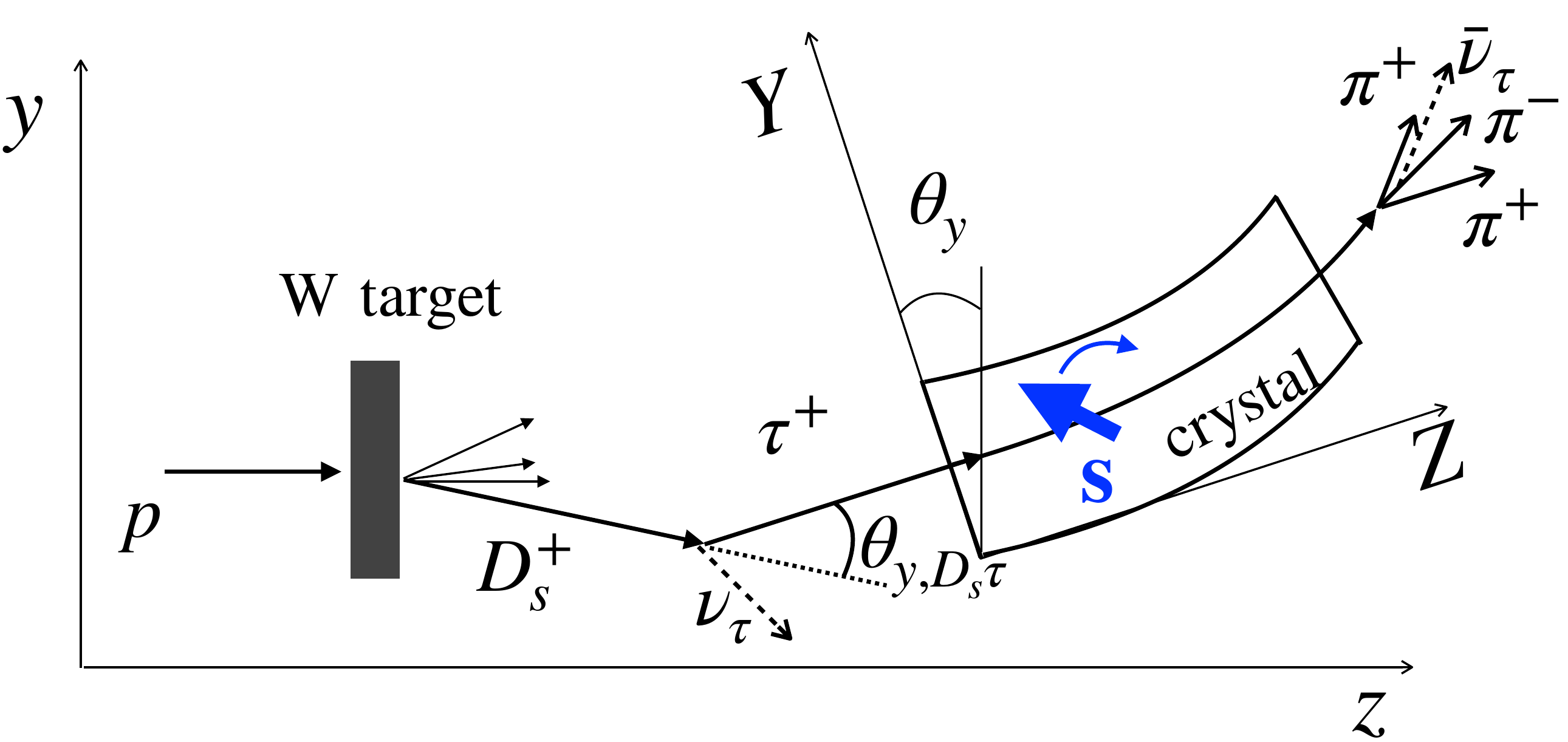} 
\caption{{\CHb (color online).}
Sketch of the fixed-target setup along with the \taup production and decay processes (not to scale). The crystal frame (\xaxis,\yaxis,\zaxis) is tilted
in the laboratory frame ($x,y,z$) by \thetay to avoid channeling of non-interacting protons.} 
\label{fig:PolCrystal}
\end{figure}

The proposed solution to provide direct measurements of the \Ptau dipole moments,
illustrated in Fig.~\ref{fig:PolCrystal},
is based on the
large production cross section of high-energy
polarized
\taup leptons,
{\CHa originating} in proton fixed-target collisions at the~\lhc.
The \taupipipi ($3\pi\neutb$) decay is considered.
A bent crystal is employed to exploit the channeling phenomenon of positively-charged particles
aligned with the crystal atomic planes within {\CHa a} few \murad.
Angular momentum conservation selects negative helicity \taup leptons in the \Dsp rest frame.
The
\taup leptons emitted at relatively large \thetayDsTau angles  with respect
 to the \Dsp
flight direction in the $yz$ plane
show enhanced polarization along the $Y$ axis, perpendicular to the crystal plane.
The Lorentz boost, making larger acceptance for forward- than for backward-emitted \taup\ {\CHb leptons},
induces a
polarization of approximately
{\CHc $\beta^\star/\beta\approx 10\%$ anti-aligned with}
the crystal $Z$ axis, where $\beta$ ($\beta^\star$) is the velocity of the \Dsp (\taup) in the laboratory (\Dsp rest) frame.
Thus,
the selection of the highest momentum candidates
enhances
the $Z$ polarization.
% thanks to the increase of the population of forward emitted \taup.
%
% is induced by the interaction of the magnetic 
% and the electric dipole moment, respectively,
% with the electric field felt by the lepton
% inside the bent crystal.
The MDM (EDM) signature is given by the spin rotation in the \yaxis\zaxis bending plane (appearance of a spin component along the \xaxis axis)
induced by the interaction with the crystal electric field.
A novel analysis technique based on multivariate classifiers
is employed to determine the rotation of the spin-polarization vector.

The vast majority of \taup leptons produced in proton fixed-target collisions
at $\sqrt{s}\approx 115$\gev comes from $\Dsp\to\taup\nu_\tau$ decays.
The corresponding production cross section 
$\sigma[pp\to\Dsp(\to\taup\neut)X]\approx1.96$\mub is estimated using
the rescaled charm production
cross section measured by the \lhcb experiment in proton-helium collisions at $\sqrt{s} = 86.6\gev$~\cite{Aaij:2018ogq},
the \cquark quark to \Dsp fragmentation fraction~\cite{Lisovyi:2015uqa,Gladilin:2014tba},
and the $\Dsp\to\taup\nu_\tau$ branching fraction~\cite{Patrignani:2016xqp}.
The conversion factor for a 7~\tev proton on a $T = 1\cm$ thick tungsten (\W) target
to produce a $\taup\to3\pi\neutb$ final state is estimated
\begin{eqnarray}
\sigma[\proton\proton\to\Dsp(\to\taup\neut) X] N_A \frac{\rho T A_N }{A_T} {\cal B}(\taup\to 3\pi\neutb)\nonumber \\
\approx 2.1\times 10^{-6}, 
\end{eqnarray}
where
$N_A$ is the Avogadro number, 
$\rho$ the target density,
$A_T$ ($A_N$) its atomic mass (mass number),
and ${\cal B}(\taup\to 3\pi\neutb)$ the \taup branching fraction~\cite{Patrignani:2016xqp}.

In a reference frame defined by the crystal edges and comoving with the channeled particle,
the \taup initial polarization ${\mathbf s}_0$ is given by the unit vector along the \Dsp momentum
in the \taup rest frame~\cite{Halzen:1984mc,Berestetsky:1982aq},
\begin{equation}
{\mathbf s}_0 = \frac{1}{\omega} \left( m_{\Ptau}{\mathbf q} - q_0 {\mathbf p} + \frac{{\mathbf q} \cdot {\mathbf p}}{p_0+m_{\Ptau}} {\mathbf p} \right),
\label{eq:s}
\end{equation}
where ${\mathbf p}$ (${\mathbf q}$) is the momentum of the \taup (\Dsp) and $p_0$ ($q_0$) its energy in the laboratory frame,
$\omega=(m^2_{\Pds}-m^2_{\Ptau})/2$, and $m_{\Pds}$ is the \Dsp mass.
The projections of ${\mathbf s}_0$ along the crystal frame axes are:
\begin{eqnarray}
s_{0,\xaxis} & \approx & \frac{m_{\Ptau} \left|{\mathbf q}\right|}{\omega} \thetaxDsTau, \nonumber \\
s_{0,\yaxis} & \approx & \frac{m_{\Ptau} \left|{\mathbf q}\right|}{\omega} \thetayDsTau,  \nonumber \\
s_{0,\zaxis} & \approx & \frac{1}{\omega} \left(\left|{\mathbf q}\right| p_0 -q_0\left|{\mathbf p}\right| \right), 
\label{eq:s_proj}
\end{eqnarray}
where
\thetaxDsTau is the angle between the \Dsp and the \taup\ {\CHa momenta} in the \x\z plane.
All angles are ${\cal O}(10^{-3})$ \rad due to the highly boosted \Ds mesons
and the small \Pds-\Ptau mass difference.
Rotational invariance and the unconstrained \thetaxDsTau in the crystal
\xaxis\zaxis plane imply a zero $s_{0,\xaxis}$ average.

\begin{figure}[htb]
\centering
\begin{tabular}{cc}
\includegraphics[width=0.35\textwidth]{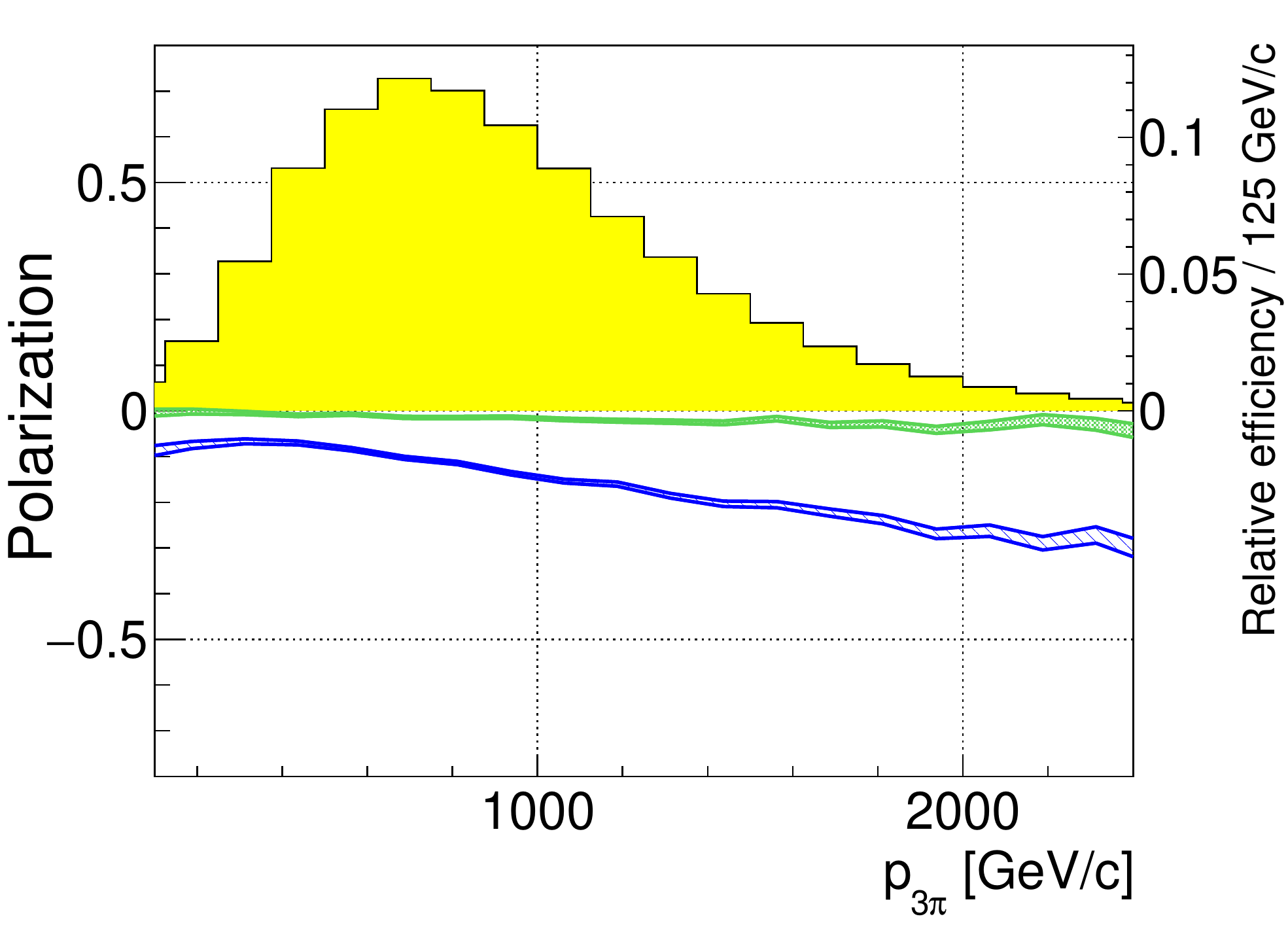}\\
\includegraphics[width=0.35\textwidth]{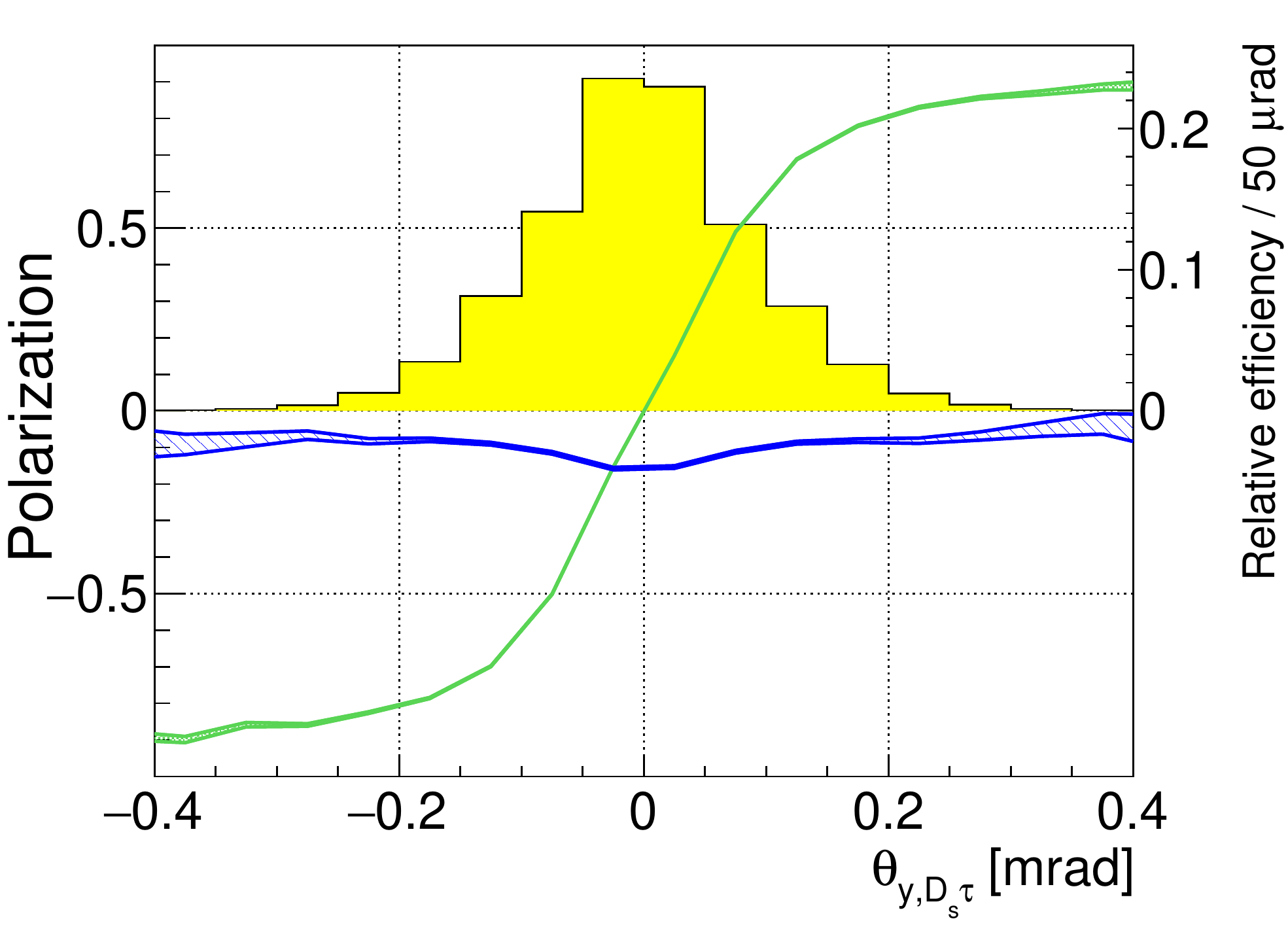}
\end{tabular}
\caption{\label{fig:p_proj} {\CHb (color online).}
  Spin-polarization projections $s_{0,\zaxis}$ ({\CHb hatched blue bands}) and $s_{0,\yaxis}$ ({\CHb solid} green)
  as a function of
  %the $3\pi$ momentum
  {\CHb $\phad$}
  (top) and \thetayDsTau (bottom).
  The histograms, normalized to unity, show the spectra of channeled \taup.
  {\CHb The bands} {\CHa represent one-standard deviation regions.}
  %arising from the limited simulation statistics.
  %The polarization along \xaxis, not shown, is consistent with zero.
  %(relative efficiency).
}
\end{figure}

Very large samples of fixed-target $\Dsp\to\taup\neut$ events
{\CHb are} produced using \pythia~\cite{Sjostrand:2006za}, \evtgen~\cite{Lange:2001uf}, and a fast simulation that generates phase-space kinematics.
The \taup channeling {\CHb is} simulated using the parameterization and procedures
described in Refs.~\cite{Biryukov1997,Bagli:2017foe}.
A
polarized sample {\CHb can be} obtained by selecting channeled \taup and imposing kinematic
requirements,
as illustrated in Fig.~\ref{fig:p_proj} for the
{\CHb optimal experimental}
layout described later.
For example, by requiring the $3\pi$ system momentum, {\CHb $\phad$,}
to exceed 1\tev a $s_{0,\zaxis}$ polarization of about $-20\%$ or higher is achieved.
Instead, selecting regions of positive or negative \thetayDsTau angles, in the following referred to as {\CHb \thetay-tagging},
a large $s_{0,\yaxis}$ polarization can be obtained.

{\CHa The interaction of the MDM (EDM) of a relativistic
charged particle channeled in a bent crystal induces spin
precession~\cite{Botella:2016ksl, Bagli:2017foe} in the bending plane
(perpendicular to the bending plane).
By measuring the spin-polarization components $s_\yaxis$ and $s_\zaxis$
($s_\xaxis$ component), it is possible to extract the MDM (EDM) information.
In particular, the appearance of {\CHb an} $s_\xaxis$ component represents the EDM signature.}
%
%
% and proved to hold for non-harmonic potentials in~\cite{Bagli:2017foe}.
%
%
The spin-polarization projections after precession in the crystal read:
%
%
%                     \approx - s_{0,z} \frac{\dpt}{\apdt}\sin\Phi , \nonumber \\  
%                     \approx s_{0,z} \frac{\apt}{\apdt}\sin\Phi , \nonumber \\
%                     \approx s_{0,z} \cos\Phi ,
%
%
%                     \approx s_{0,y} \frac{\dpt\apt}{\apdt^2} \left( 1-\cos\Phi \right) , \nonumber \\  
%                     \approx s_{0,y} \left(\frac{\dpt^2}{\apdt^2} + \frac{\apt^2}{\apdt^2} \cos\Phi\right), \nonumber \\
%                     \approx - s_{0,y} \frac{\apt}{\apdt} \sin\Phi .
%
%
\begin{align}
s_\xaxis & 
                     \approx -s_{0,\zaxis}  \frac{\dpt}{\apdt}\sin\Phi  + s_{0,\yaxis} \frac{\dpt\apt}{\apdt^2} \left( 1-\cos\Phi \right) , \nonumber \\  
s_\yaxis & 
                     \approx s_{0,\zaxis} \frac{\apt}{\apdt}\sin\Phi    + s_{0,\yaxis} \left(\frac{\dpt^2}{\apdt^2} + \frac{\apt^2}{\apdt^2} \cos\Phi\right), \nonumber \\
s_\zaxis & 
                     \approx s_{0,\zaxis} \cos\Phi                      - s_{0,\yaxis} \frac{\apt}{\apdt} \sin\Phi ,
\label{eq:eom_main_szsy}
\end{align}
where
$\apt = \at+\frac{1}{1+\gamma}$,
$\dpt = \dt/2$,
$\apdt = \sqrt{\apt^2+\dpt^2}$,
{\CHa and $\Phi = \gamma \theta_C \apdt$ is the precession angle,
which is proportional to the \taup Lorentz factor $\gamma$ and 
 the crystal bending angle $\theta_C$.}
Equation~(\ref{eq:eom_main_szsy}) holds at ${\cal O}(10^{-2})$ precision, while expressions
at ${\cal O}(10^{-5})$ are reported in the supplemental material~\cite{supplemental}.

{\CHa
A
technique based on multivariate classifiers is explored to extract the \taup polarization vector
without prior knowledge of the
detailed
decay dynamics and of the \taup energy.}
A classifier discriminating between \taup
with full positive ($+1$) and negative ($-1$) polarization along each crystal frame axis is built.
The classifiers are trained on  
{\CHb simulated events}
{\CHb and are based} upon
variables describing the decay distribution.
The used variables
{\CHa that provide sensitivity to the \taup spin polarization,}
referred to with the symbol ${\bm \zeta}$, are:
the angles between the ${3\pi}$ momentum in the \taup rest frame and the crystal frame axes,
the angles describing the $3\pi$ decay plane orientation in the $3\pi$ rest frame with respect to the crystal frame axes,
{\CHb and two- and three-pion invariant masses.}
The \taup\ {\CHb momentum} is estimated by applying {\CHb kinematic} corrections, determined from simulated events,
to the measured
{\CHb \phad} vector as a function of its magnitude and direction.
In absence of the \taup production vertex, the flight direction is assumed to be that connecting the \Dsp production
vertex and the \taup decay vertex, lying in the crystal channeling plane. The vertex positions are smeared according to Gaussian distributions to mimic experimental
resolutions, assumed to be 13\mum (70\mum) for the production vertex in the longitudinal
(transverse) direction with respect to the beam, and 
100\mum (1\mm) for the decay vertex.

{\CHa The polarization component $s_i$ {\CHb along} the $i$-th crystal frame axis {\CHb $(i =\xaxis,\yaxis,\zaxis$)}
is extracted by fitting the classifier distribution \Wieta on data, 
\begin{align}
%       &= \frac{\Wipeta+\Wimeta}{2} + s_i \frac{\Wipeta-\Wimeta}{2}.
\Wieta &= \frac{1}{2} \left[ (1+s_i) \Wipeta + (1-s_i)  \Wimeta \right] ,
\label{eq:pol_template}
\end{align}
where $\eta \equiv \eta({\bm \zeta})$ is the classifier response,
and \Wipmeta the templates representing the response for $\pm 1$ polarizations.}

The statistical separation between templates also represents the squared average event {\CHb information~\cite{Kendall:1983}}
of the polarization (at $s_i=0$)~\cite{Davier:1992nw},
\begin{equation}
S_i^2 = \frac{1}{N_\taup^{\rm rec} \sigma_i^2} = \left\langle \left( \frac{\Wipeta-\Wimeta}{ \Wipeta+\Wimeta } \right)^2 \right\rangle ,
\end{equation}
where $\sigma_i$ is the uncertainty on $s_i$,
{\CHb and $N_\taup^{\rm rec}$ is the number of channeled and reconstructed \taup.}
The template fit results for $s_{\yaxis}$ polarization are shown in Fig.~\ref{fig:template_fit},
while {\CHb those} for $s_{\xaxis}$ and $s_{\zaxis}$
are shown in the supplemental material~\cite{supplemental}.
The estimated average event information is $S_\xaxis\approx S_\yaxis\approx0.42$ and $S_\zaxis\approx 0.23$,
using either Multilayer Perceptron Networks or Boosted Decision Trees~\cite{Hocker:2007ht},
to be compared to the ideal value of $0.58$ reached
in case the complete kinematics of the \taup decay is reconstructed~\cite{Davier:1992nw}. The difficulty in determining
the \taup momentum, due to the {\CHb undetected} \neutb,
affects mainly the determination of the $s_{\zaxis}$ polarization.

% moved here to avoid floating of the figure to the following page
\begin{figure}[htb]
\centering
\begin{tabular}{cc}
\includegraphics[width=0.35\textwidth]{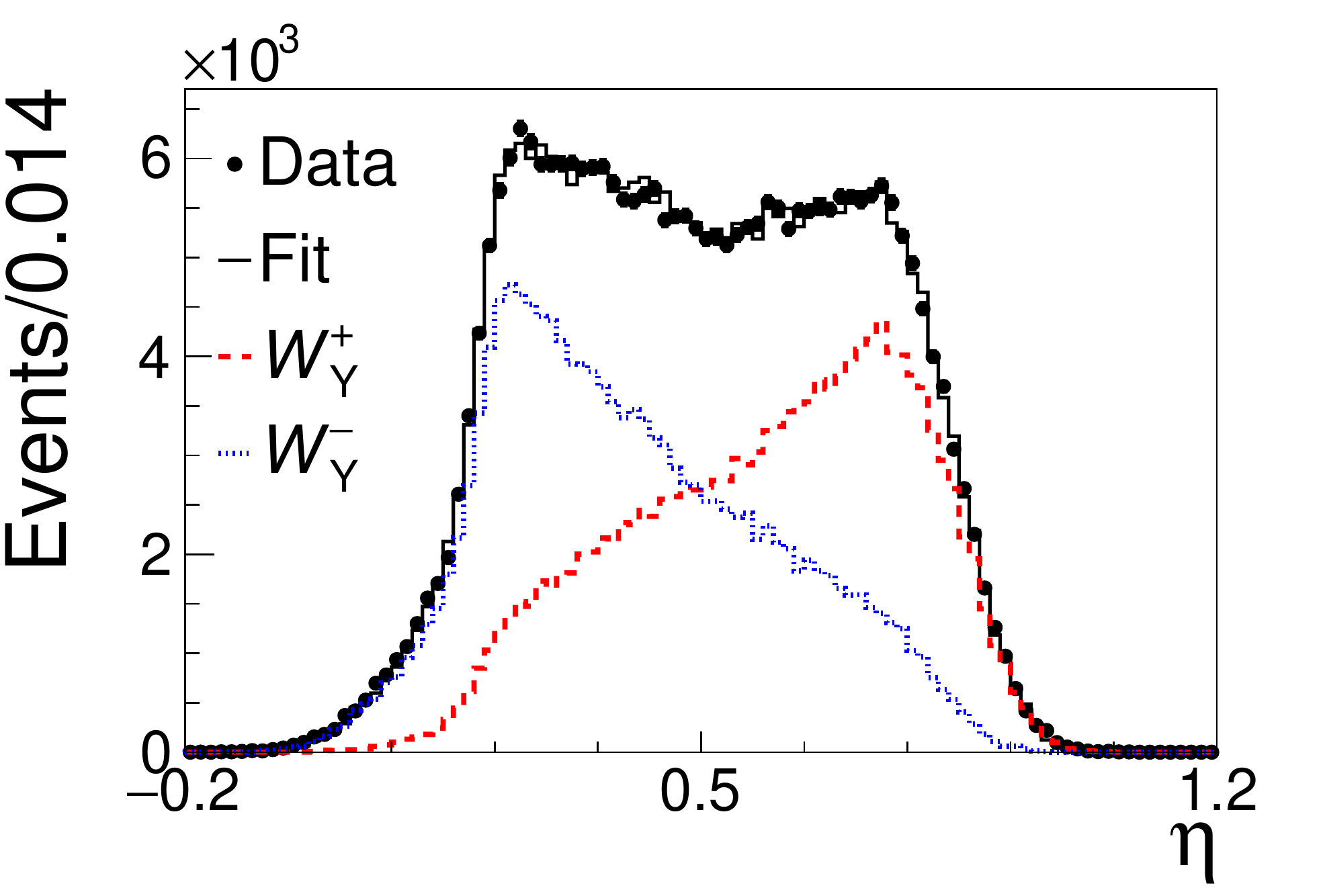} \\
\end{tabular}
\caption{\label{fig:template_fit} {\CHb (color online).}
Template fit results to the classifier response $\eta$ using simulated events for $s_{\yaxis}$ polarization.
The normalized separation between $\Wypeta$ ({\CHb dashed red line}) and $\Wymeta$ ({\CHb dotted blue line}) determines the average
event information $S_\yaxis\approx 0.42$.
}
\end{figure}

% with ${\cal O}(10^{-2})$ approximation.
For small $\Phi$ (as $\gamma\theta_C \sim 10$ and $\apdt\sim 10^{-3}$) and $s_{0,\zaxis}$ initial polarization,
the statistical uncertainties on $a$ and $d$ are estimated from Eq.~(\ref{eq:eom_main_szsy}) as
\begin{eqnarray}
\sigma_{a} \approx \frac{1}{{\CHa S_\yaxis} s_{0,\zaxis} \gamma \theta_C } \frac{1}{\sqrt{N_\taup^{\rm rec}}} , \ \ \
\sigma_{d} \approx \frac{2}{{\CHa S_\xaxis} s_{0,\zaxis} \gamma \theta_C } \frac{1}{\sqrt{N_\taup^{\rm rec}}} .
\label{eq:sz_uncertainty}
\end{eqnarray}
{\CHb For} $s_{0,\yaxis}$ initial polarization,
\begin{eqnarray}
\sigma_{a} \approx \frac{1}{{\CHa S_\zaxis} s_{0,\yaxis} \gamma \theta_C } \frac{1}{\sqrt{N_\taup^{\rm rec}}} , \ \ \
\sigma_{d} \approx \frac{2}{{\CHa S_\xaxis} s_{0,\yaxis} (\gamma \theta_C)^2 \apt } \frac{1}{\sqrt{N_\taup^{\rm rec}}} ,
\label{eq:sy_uncertainty}
\end{eqnarray}
which show comparable sensitivity to $a$ but disfavored  by a factor
$1/(\gamma\theta_C\apt)\sim 100$  to $d$ with respect to Eq.~(\ref{eq:sz_uncertainty})
for initial $s_{0,\zaxis}$ polarization.

{\CHb The} optimization of the
{\CHb experimental} layout
{\CHb is}
performed
{\CHb using simulated events}
for {\CHb the case of} initial $s_{0,\zaxis}$ polarization.
{\CHb The region} of minimal uncertainty on $a$ and $d$
{\CHb is determined}
using a
scan
in the {\CHb $(\thc,\Lc,\thetay,\Ltarc)$}
parameter space,
where
\thc (\Lc) is the crystal bending angle (length)
{\CHb and}
\Ltarc the distance between the target and {\CHb crystal.}
{\CHb Channeled \taup are required to have $\phad>800\gevc$ to enhance $s_{0,\zaxis}$ polarization, and}
{\CHa to originate before the crystal and to decay after the crystal}
{\CHb to insure maximum $\Phi$ precession angle.}
For a \Ge (\Si) crystal tilted by $\thetay=0.1\mrad$, the optimal parameters 
$\thc \approx 16\mrad$,
$\Lc \approx 8~(11)\cm$,
and $\Ltarc\approx 12\cm$
%and $\phad>800\gevc$
are obtained
(see supplemental material~\cite{supplemental}).
{\CHa
The \Ge crystal provides relatively high channeling efficiency,
$\approx 6.3\times 10^{-6}$, a factor {\CHb of} three higher than for \Si.
Recently, crystal prototypes
with similar length and bending angle have been tested on beam at the CERN SPS~\cite{Mazzolari:2018}.}
%
%  Regions of minimal uncertainty for \at and \dt as a function of the crystal parameters \Lc and \thc for Ge (red) and Si (blue),
%  for initial \zaxis polarization.
%  The lines represent regions whose uncertainties on \at and \dt are increased by 10\% with respect to the minimum (point).
%
%
{\CHb The selected \taup sample}
{\CHa has $s_{0,\zaxis}\approx -18\%$,
$s_{0,\yaxis}\approx 0\%$ {\CHb polarization,} and average Lorentz factor $\gamma\approx 800$.}
A $s_{0,\yaxis}\approx\mp40\%$ polarization can be achieved with a \thetay-tagging that discriminates between positive and negative \thetayDsTau angles.
%
%
%
%                             & \Ge & \Ge              &  \Si & \Si \\
%                             &     & \thetay-tagging  &      & \thetay-tagging \\
%
%
Information statistically correlated with \thetayDsTau is required for \thetay-tagging.
A possible strategy could be the exploitation of the global event topology,
\eg kinematic distributions of particles associated with the interaction point where the \Dsp
is produced. The relatively large separation between the target and the crystal
would allow
{\CHb for}
additional instrumentation, \eg several layers of pixel
{\CHb radiation-hard}
diamond sensors could be used to reconstruct the \Dsp trajectory. 
Another possibility would be to place
a second bent crystal to channel the \Dsp
using a layout similar to that suggested in Ref.~\cite{Fomin:2018ybj},
inducing $s_{0,\yaxis}\approx \mp60\%$ for tagged events with an efficiency of {\CHb a few percent.}
%
%
%
% MOVED TO THE END OF THE ARTICLE WHEN SYSTEMATIC UNCERTAINTIES ARE DISCUSSED
%
%

%
%

%
%
%
%
%
%

\begin{figure}[htb]
\centering
\begin{tabular}{cc}
\includegraphics[width=0.40\textwidth]{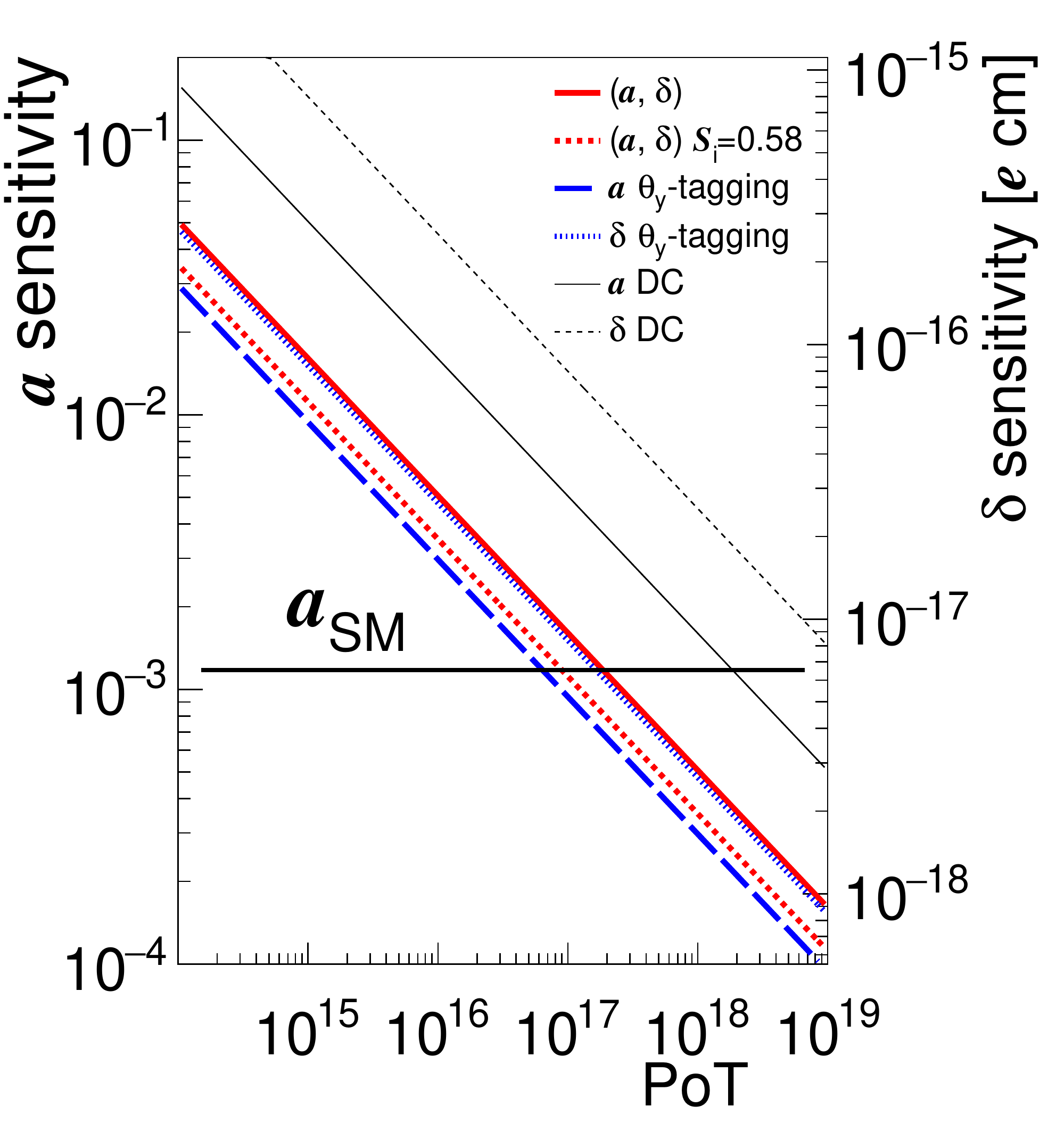}
\end{tabular}
\caption{\label{fig:sensitivity} (color online).
Estimated sensitivities for \at and \deltat as a function of the number of protons on
{\CHc a 2.5 cm thick W}
target (\pot) for a \Ge crystal with optimal parameters {\CHb (thick solid red line)},
compared to other configurations (see text).
{\CHb These are labeled as (\at, \deltat) when the corresponding lines overlap.}
The SM model prediction for $a$~\cite{Eidelman:2007sb} is also indicated.
}
\end{figure}

{\CHb Dipole moment sensitivities}
are assessed from a large number of pseudoexperiments generated and fit using a probability density function based on the spin precession
equation of motion reported in Eq.~(\ref{eq:eom_main_szsy}), and
the classifier distributions in Eq.~(\ref{eq:pol_template}).
Figure~\ref{fig:sensitivity} illustrates the estimated sensitivities
as a function of the number of impinging protons for a \Ge crystal with optimal parameters {\CHb (thick solid red line)}.
Sensitivities for other configurations with maximum average event information {\CHa $S_i=0.58$} {\CHb (thick dotted red line)},
\thetay-tagging based on a discrimination between positive and negative \thetayDsTau\ {\CHb with} ideal
{\CHb tagging} efficiency of 100\% {\CHb (thick dashed and hatched blue lines)},
and the double crystal (DC) option proposed in Ref.~\cite{Fomin:2018ybj} {\CHb (thin solid and dotted black lines)}, are also shown for comparison.
A detector reconstruction efficiency of 40\% is assumed.
The corresponding sensitivities for \Si are about a factor two worse. 
%
% with similar kinematic properties to the signal.
% $s_{0,\zaxis}$ and $s_{0,\yaxis}$ can be accurately determined from the ${\bf s}$ vector of channeled \taup and
%
%

%
%

%
%
% their common vertex, which identifies the decay position.
% to be located after the crystal, at a distance  $\Lc+\Ltarc\gsim 20\cm$ from the interaction point, 
%  the non channeled particles are suppressed at negligibly small level.
%  vetoed using the reconstructed invariant mass and the particle identification information.

The channeling process keeps the high momentum unchanged while deflecting the \taup
at the bending angle $\theta_C \approx 16\mrad$.
This signature can be identified in the $3\pi\neutb$ decays through
the reconstruction of the $3\pi$ vertex and momentum.
For highly-boosted particles with $\gamma\approx800$ the latter defines
the \taup direction with an uncertainty of $\approx0.5\mrad$, mainly due to the missing
\neutb.
The contribution of non-channeled
{\CHb leptons}
is reduced to a negligible level $<0.3\%$
using the following selection criteria:
$\phad \ge 800\gevc$, $3\pi$ momentum direction consistent with
$\theta_C$ within 1.5\mrad, and the $3\pi$ vertex located after the crystal,
at a distance  $\Lc+\Ltarc\gsim 20\cm$ from the interaction point.
{\CHb With these requirements, 28\% of the \taup candidates}
are channeled through a
fraction of the crystal length. These are mainly events in which the \Dsp decays inside the crystal
or the $\taup$  does not reach the end of the crystal, either because
it decays or is dechanneled.
Nevertheless,
only \taup particles that travel almost through the entire crystal are {\CHb selected.}
{\CHb They} experience {\CHb a} very similar electromagnetic field,
inducing a relatively small bias on the spin precession angle $\Phi$ of $1.4\%$
that can be corrected.
Background contributions from channeled hadron decays with $3\pi$ in the final state,
\eg~\Dp, \Dsp mesons, \Lcp baryons, can be vetoed using the reconstructed invariant mass
and event information from a dedicated detecting apparatus.
{\CHb Systematic effects} {\CHa could arise from the limited knowledge of the crystal position and orientation,} {\CHb the initial polarization, and the
\taup momentum.} {\CHa Those uncertainties} {\CHb can be controlled using up- and down-bending crystals, inducing opposite spin precession~\cite{Bagli:2017foe},
by reconstructing unchanneled \taup decays with kinematic properties similar to the signal, and by using detailed simulations of the experimental setup
calibrated with data.}
Possible effects due to \taup weak interactions with the crystal are estimated to be negligible~\cite{Baryshevsky:2018dqq} compared to the
sensitivity and can be removed by using different crystal bending orientations.

In summary, a novel method for the direct measurement of the \Ptau MDM and EDM
has been
presented with interesting perspective for a stringent test of the SM and search of new physics.
The fixed-target setup and the analysis technique
have been discussed along
with sensitivity projections for possible future scenarios. 
The SM prediction for the \Ptau MDM could be verified experimentally
with
{\CHb a sample of}
around
{\CHc $10^{17}$\pot,}
whereas at the same time a search for the \Ptau EDM
at the level of $10^{-17}~e\cm$ or below
could be performed.
{\CHc This would require about 10\% of the protons storaged during a decade of \lhc operation~\cite{ApollinariG.:2017ojx}.}
In preparation of a possible future {\CHc experiment}
this method could be
tested
using the fixed-target setup
proposed for the study of heavy baryons~\cite{Botella:2016ksl,Bezshyyko:2017var,Bagli:2017foe} with 
the \lhcb apparatus.
The possibility of a test or an
experiment at the CERN SPS will also be explored.

\indent
We express our gratitude for stimulating discussions to V.~Baryshevsky, F.J.~Botella, G.~Cavoto, A.~S.~Fomin, A.~Mazzolari, A.~Pich and J.~Walsh.
We acknowledge support from INFN (Italy), MINECO and GVA (Spain), the Severo Ochoa excellence certification SEV-2014-0398-01,
and the ERC Consolidator Grant SELDOM G.A. 771642.

\bibliography{biblio}% Produces the bibliography via BibTeX.

\begin{thebibliography}{43}
\expandafter\ifx\csname natexlab\endcsname\relax\def\natexlab#1{#1}\fi
\expandafter\ifx\csname bibnamefont\endcsname\relax
  \def\bibnamefont#1{#1}\fi
\expandafter\ifx\csname bibfnamefont\endcsname\relax
  \def\bibfnamefont#1{#1}\fi
\expandafter\ifx\csname citenamefont\endcsname\relax
  \def\citenamefont#1{#1}\fi
\expandafter\ifx\csname url\endcsname\relax
  \def\url#1{\texttt{#1}}\fi
\expandafter\ifx\csname urlprefix\endcsname\relax\def\urlprefix{URL }\fi
\providecommand{\bibinfo}[2]{#2}
\providecommand{\eprint}[2][]{\url{#2}}

\bibitem[{\citenamefont{Mohr et~al.}(2016)\citenamefont{Mohr, Newell, and
  Taylor}}]{Mohr:2015ccw}
\bibinfo{author}{\bibfnamefont{P.~J.} \bibnamefont{Mohr}},
  \bibinfo{author}{\bibfnamefont{D.~B.} \bibnamefont{Newell}},
  \bibnamefont{and} \bibinfo{author}{\bibfnamefont{B.~N.}
  \bibnamefont{Taylor}}, \bibinfo{journal}{Rev. Mod. Phys.}
  \textbf{\bibinfo{volume}{88}}, \bibinfo{pages}{035009}
  (\bibinfo{year}{2016}).

\bibitem[{\citenamefont{Andreev et~al.}(2018)}]{Andreev:2018ayy}
\bibinfo{author}{\bibfnamefont{V.}~\bibnamefont{Andreev}} \bibnamefont{et~al.}
  (\bibinfo{collaboration}{ACME collaboration}), \bibinfo{journal}{Nature}
  \textbf{\bibinfo{volume}{562}}, \bibinfo{pages}{355} (\bibinfo{year}{2018}).

\bibitem[{\citenamefont{Bennett et~al.}(2006)}]{Bennett:2006fi}
\bibinfo{author}{\bibfnamefont{G.~W.} \bibnamefont{Bennett}}
  \bibnamefont{et~al.} (\bibinfo{collaboration}{Muon (g-2) collaboration}),
  \bibinfo{journal}{Phys. Rev.} \textbf{\bibinfo{volume}{D73}},
  \bibinfo{pages}{072003} (\bibinfo{year}{2006}).

\bibitem[{\citenamefont{Bennett et~al.}(2009)}]{Bennett:2008dy}
\bibinfo{author}{\bibfnamefont{G.~W.} \bibnamefont{Bennett}}
  \bibnamefont{et~al.} (\bibinfo{collaboration}{Muon (g-2) collaboration}),
  \bibinfo{journal}{Phys. Rev.} \textbf{\bibinfo{volume}{D80}},
  \bibinfo{pages}{052008} (\bibinfo{year}{2009}).

\bibitem[{\citenamefont{Pendlebury et~al.}(2015)}]{Afach:2015sja}
\bibinfo{author}{\bibfnamefont{J.~M.} \bibnamefont{Pendlebury}}
  \bibnamefont{et~al.}, \bibinfo{journal}{Phys. Rev.}
  \textbf{\bibinfo{volume}{D92}}, \bibinfo{pages}{092003}
  (\bibinfo{year}{2015}).

\bibitem[{\citenamefont{Schneider et~al.}(2017)}]{Schneider:2017}
\bibinfo{author}{\bibfnamefont{G.}~\bibnamefont{Schneider}}
  \bibnamefont{et~al.}, \bibinfo{journal}{Science 24}
  \textbf{\bibinfo{volume}{358}}, \bibinfo{pages}{1081} (\bibinfo{year}{2017}).

\bibitem[{\citenamefont{Sahoo}(2017)}]{Sahoo:2016zvr}
\bibinfo{author}{\bibfnamefont{B.}~\bibnamefont{Sahoo}},
  \bibinfo{journal}{Phys. Rev.} \textbf{\bibinfo{volume}{D95}},
  \bibinfo{pages}{013002} (\bibinfo{year}{2017}).

\bibitem[{\citenamefont{Graner et~al.}(2016)\citenamefont{Graner, Chen,
  Lindahl, and Heckel}}]{Graner:2016ses}
\bibinfo{author}{\bibfnamefont{B.}~\bibnamefont{Graner}},
  \bibinfo{author}{\bibfnamefont{Y.}~\bibnamefont{Chen}},
  \bibinfo{author}{\bibfnamefont{E.~G.} \bibnamefont{Lindahl}},
  \bibnamefont{and} \bibinfo{author}{\bibfnamefont{B.~R.}
  \bibnamefont{Heckel}}, \bibinfo{journal}{Phys. Rev. Lett.}
  \textbf{\bibinfo{volume}{116}}, \bibinfo{pages}{161601}
  (\bibinfo{year}{2016}), \bibinfo{note}{[Erratum: Phys. Rev. Lett.119, 119901
  (2017)]}.

\bibitem[{\citenamefont{Baryshevsky}(2016)}]{Baryshevsky:2016cul}
\bibinfo{author}{\bibfnamefont{V.~G.} \bibnamefont{Baryshevsky}},
  \bibinfo{journal}{Phys. Lett.} \textbf{\bibinfo{volume}{B757}},
  \bibinfo{pages}{426} (\bibinfo{year}{2016}).

\bibitem[{\citenamefont{Botella et~al.}(2017)\citenamefont{Botella,
  Garcia~Martin, Marangotto, Martinez~Vidal, Merli, Neri, Oyanguren, and
  Ruiz~Vidal}}]{Botella:2016ksl}
\bibinfo{author}{\bibfnamefont{F.~J.} \bibnamefont{Botella}},
  \bibinfo{author}{\bibfnamefont{L.~M.} \bibnamefont{Garcia~Martin}},
  \bibinfo{author}{\bibfnamefont{D.}~\bibnamefont{Marangotto}},
  \bibinfo{author}{\bibfnamefont{F.}~\bibnamefont{Martinez~Vidal}},
  \bibinfo{author}{\bibfnamefont{A.}~\bibnamefont{Merli}},
  \bibinfo{author}{\bibfnamefont{N.}~\bibnamefont{Neri}},
  \bibinfo{author}{\bibfnamefont{A.}~\bibnamefont{Oyanguren}},
  \bibnamefont{and}
  \bibinfo{author}{\bibfnamefont{J.}~\bibnamefont{Ruiz~Vidal}},
  \bibinfo{journal}{Eur. Phys. J.} \textbf{\bibinfo{volume}{C77}},
  \bibinfo{pages}{181} (\bibinfo{year}{2017}).

\bibitem[{\citenamefont{Baryshevsky}(2017)}]{Baryshevsky:2017yhk}
\bibinfo{author}{\bibfnamefont{V.~G.} \bibnamefont{Baryshevsky}},
  \bibinfo{journal}{Nucl. Instrum. Meth.} \textbf{\bibinfo{volume}{B402}},
  \bibinfo{pages}{5} (\bibinfo{year}{2017}).

\bibitem[{\citenamefont{Fomin et~al.}(2017)\citenamefont{Fomin, Korchin,
  Stocchi, Bezshyyko, Burmistrov, Fomin, Kirillin, Massacrier, Natochii, Robbe
  et~al.}}]{Bezshyyko:2017var}
\bibinfo{author}{\bibfnamefont{A.~S.} \bibnamefont{Fomin}},
  \bibinfo{author}{\bibfnamefont{A.~Y.} \bibnamefont{Korchin}},
  \bibinfo{author}{\bibfnamefont{A.}~\bibnamefont{Stocchi}},
  \bibinfo{author}{\bibfnamefont{O.~A.} \bibnamefont{Bezshyyko}},
  \bibinfo{author}{\bibfnamefont{L.}~\bibnamefont{Burmistrov}},
  \bibinfo{author}{\bibfnamefont{S.~P.} \bibnamefont{Fomin}},
  \bibinfo{author}{\bibfnamefont{I.~V.} \bibnamefont{Kirillin}},
  \bibinfo{author}{\bibfnamefont{L.}~\bibnamefont{Massacrier}},
  \bibinfo{author}{\bibfnamefont{A.}~\bibnamefont{Natochii}},
  \bibinfo{author}{\bibfnamefont{P.}~\bibnamefont{Robbe}},
  \bibnamefont{et~al.}, \bibinfo{journal}{JHEP} \textbf{\bibinfo{volume}{08}},
  \bibinfo{pages}{120} (\bibinfo{year}{2017}).

\bibitem[{\citenamefont{Bagli et~al.}(2017)\citenamefont{Bagli, Bandiera,
  Cavoto, Guidi, Henry, Marangotto, Martinez~Vidal, Mazzolari, Merli, Neri
  et~al.}}]{Bagli:2017foe}
\bibinfo{author}{\bibfnamefont{E.}~\bibnamefont{Bagli}},
  \bibinfo{author}{\bibfnamefont{L.}~\bibnamefont{Bandiera}},
  \bibinfo{author}{\bibfnamefont{G.}~\bibnamefont{Cavoto}},
  \bibinfo{author}{\bibfnamefont{V.}~\bibnamefont{Guidi}},
  \bibinfo{author}{\bibfnamefont{L.}~\bibnamefont{Henry}},
  \bibinfo{author}{\bibfnamefont{D.}~\bibnamefont{Marangotto}},
  \bibinfo{author}{\bibfnamefont{F.}~\bibnamefont{Martinez~Vidal}},
  \bibinfo{author}{\bibfnamefont{A.}~\bibnamefont{Mazzolari}},
  \bibinfo{author}{\bibfnamefont{A.}~\bibnamefont{Merli}},
  \bibinfo{author}{\bibfnamefont{N.}~\bibnamefont{Neri}}, \bibnamefont{et~al.},
  \bibinfo{journal}{Eur. Phys. J.} \textbf{\bibinfo{volume}{C77}},
  \bibinfo{pages}{828} (\bibinfo{year}{2017}).

\bibitem[{\citenamefont{Baryshevsky}(2018)}]{Baryshevsky:2018dqq}
\bibinfo{author}{\bibfnamefont{V.~G.} \bibnamefont{Baryshevsky}}
  (\bibinfo{year}{2018}), \eprint{arXiv:1803.05770}.

\bibitem[{\citenamefont{Samuel et~al.}(1991)\citenamefont{Samuel, Li, and
  Mendel}}]{Samuel:1990su}
\bibinfo{author}{\bibfnamefont{M.~A.} \bibnamefont{Samuel}},
  \bibinfo{author}{\bibfnamefont{G.-w.} \bibnamefont{Li}}, \bibnamefont{and}
  \bibinfo{author}{\bibfnamefont{R.}~\bibnamefont{Mendel}},
  \bibinfo{journal}{Phys. Rev. Lett.} \textbf{\bibinfo{volume}{67}},
  \bibinfo{pages}{668} (\bibinfo{year}{1991}), \bibinfo{note}{[Erratum: Phys.
  Rev. Lett.69,995(1992)]}.

\bibitem[{\citenamefont{Fomin et~al.}(2019)\citenamefont{Fomin, Korchin,
  Stocchi, Barsuk, and Robbe}}]{Fomin:2018ybj}
\bibinfo{author}{\bibfnamefont{A.~S.} \bibnamefont{Fomin}},
  \bibinfo{author}{\bibfnamefont{A.~Y.} \bibnamefont{Korchin}},
  \bibinfo{author}{\bibfnamefont{A.}~\bibnamefont{Stocchi}},
  \bibinfo{author}{\bibfnamefont{S.}~\bibnamefont{Barsuk}}, \bibnamefont{and}
  \bibinfo{author}{\bibfnamefont{P.}~\bibnamefont{Robbe}},
  \bibinfo{journal}{{\CHb JHEP}} \textbf{\bibinfo{volume}{03}},
  \bibinfo{pages}{156} (\bibinfo{year}{2019}), \eprint{1810.06699}.

\bibitem[{\citenamefont{Leader}(2011)}]{Leader2011}
\bibinfo{author}{\bibfnamefont{E.}~\bibnamefont{Leader}},
  \emph{\bibinfo{title}{{Spin in particle physics}}}, vol.~\bibinfo{volume}{15}
  of \emph{\bibinfo{series}{Camb. Monogr. Part. Phys. Nucl. Phys. Cosmol.}}
  (\bibinfo{publisher}{Cambridge University Press},
  \bibinfo{address}{Cambridge}, \bibinfo{year}{2011}).

\bibitem[{\citenamefont{Eidelman and Passera}(2007)}]{Eidelman:2007sb}
\bibinfo{author}{\bibfnamefont{S.}~\bibnamefont{Eidelman}} \bibnamefont{and}
  \bibinfo{author}{\bibfnamefont{M.}~\bibnamefont{Passera}},
  \bibinfo{journal}{Mod. Phys. Lett.} \textbf{\bibinfo{volume}{A22}},
  \bibinfo{pages}{159} (\bibinfo{year}{2007}).

\bibitem[{\citenamefont{Pospelov and Ritz}(2014)}]{Pospelov:2013sca}
\bibinfo{author}{\bibfnamefont{M.}~\bibnamefont{Pospelov}} \bibnamefont{and}
  \bibinfo{author}{\bibfnamefont{A.}~\bibnamefont{Ritz}},
  \bibinfo{journal}{Phys. Rev.} \textbf{\bibinfo{volume}{D89}},
  \bibinfo{pages}{056006} (\bibinfo{year}{2014}).

\bibitem[{\citenamefont{Pich}(2014)}]{Pich:2013lsa}
\bibinfo{author}{\bibfnamefont{A.}~\bibnamefont{Pich}}, \bibinfo{journal}{Prog.
  Part. Nucl. Phys.} \textbf{\bibinfo{volume}{75}}, \bibinfo{pages}{41}
  (\bibinfo{year}{2014}), \bibinfo{note}{and references therein}.

\bibitem[{\citenamefont{Dekens et~al.}(2019)\citenamefont{Dekens, de~Vries,
  Jung, and Vos}}]{Dekens:2018bci}
\bibinfo{author}{\bibfnamefont{W.}~\bibnamefont{Dekens}},
  \bibinfo{author}{\bibfnamefont{J.}~\bibnamefont{de~Vries}},
  \bibinfo{author}{\bibfnamefont{M.}~\bibnamefont{Jung}}, \bibnamefont{and}
  \bibinfo{author}{\bibfnamefont{K.~K.} \bibnamefont{Vos}},
  \bibinfo{journal}{JHEP} \textbf{\bibinfo{volume}{01}}, \bibinfo{pages}{069}
  (\bibinfo{year}{2019}).

\bibitem[{\citenamefont{Abdallah et~al.}(2004)}]{Abdallah:2003xd}
\bibinfo{author}{\bibfnamefont{J.}~\bibnamefont{Abdallah}} \bibnamefont{et~al.}
  (\bibinfo{collaboration}{DELPHI collaboration}), \bibinfo{journal}{Eur. Phys.
  J.} \textbf{\bibinfo{volume}{C35}}, \bibinfo{pages}{159}
  (\bibinfo{year}{2004}).

\bibitem[{\citenamefont{Inami et~al.}(2003)}]{Inami:2002ah}
\bibinfo{author}{\bibfnamefont{K.}~\bibnamefont{Inami}} \bibnamefont{et~al.}
  (\bibinfo{collaboration}{Belle collaboration}), \bibinfo{journal}{Phys.
  Lett.} \textbf{\bibinfo{volume}{B551}}, \bibinfo{pages}{16}
  (\bibinfo{year}{2003}).

\bibitem[{\citenamefont{Hayreter and Valencia}(2015)}]{Hayreter:2015cia}
\bibinfo{author}{\bibfnamefont{A.}~\bibnamefont{Hayreter}} \bibnamefont{and}
  \bibinfo{author}{\bibfnamefont{G.}~\bibnamefont{Valencia}},
  \bibinfo{journal}{JHEP} \textbf{\bibinfo{volume}{07}}, \bibinfo{pages}{174}
  (\bibinfo{year}{2015}).

\bibitem[{\citenamefont{Chen and Wu}(2018)}]{Chen:2018cxt}
\bibinfo{author}{\bibfnamefont{X.}~\bibnamefont{Chen}} \bibnamefont{and}
  \bibinfo{author}{\bibfnamefont{Y.}~\bibnamefont{Wu}} (\bibinfo{year}{2018}),
  \eprint{arXiv:1803.00501}.

\bibitem[{\citenamefont{Aaij et~al.}(2018)}]{Aaij:2018ogq}
\bibinfo{author}{\bibfnamefont{R.}~\bibnamefont{Aaij}} \bibnamefont{et~al.}
  (\bibinfo{collaboration}{LHCb collaboration}) (\bibinfo{year}{2018}),
  \eprint{arXiv:1810.07907}.

\bibitem[{\citenamefont{Lisovyi et~al.}(2016)\citenamefont{Lisovyi, Verbytskyi,
  and Zenaiev}}]{Lisovyi:2015uqa}
\bibinfo{author}{\bibfnamefont{M.}~\bibnamefont{Lisovyi}},
  \bibinfo{author}{\bibfnamefont{A.}~\bibnamefont{Verbytskyi}},
  \bibnamefont{and} \bibinfo{author}{\bibfnamefont{O.}~\bibnamefont{Zenaiev}},
  \bibinfo{journal}{Eur. Phys. J.} \textbf{\bibinfo{volume}{C76}},
  \bibinfo{pages}{397} (\bibinfo{year}{2016}).

\bibitem[{\citenamefont{Gladilin}(2015)}]{Gladilin:2014tba}
\bibinfo{author}{\bibfnamefont{L.}~\bibnamefont{Gladilin}},
  \bibinfo{journal}{Eur. Phys. J.} \textbf{\bibinfo{volume}{C75}},
  \bibinfo{pages}{19} (\bibinfo{year}{2015}).

\bibitem[{\citenamefont{Patrignani et~al.}(2016)}]{Patrignani:2016xqp}
\bibinfo{author}{\bibfnamefont{C.}~\bibnamefont{Patrignani}}
  \bibnamefont{et~al.} (\bibinfo{collaboration}{Particle Data Group}),
  \bibinfo{journal}{Chin. Phys.} \textbf{\bibinfo{volume}{C40}},
  \bibinfo{pages}{100001} (\bibinfo{year}{2016}).

\bibitem[{\citenamefont{Halzen and Martin}(1984)}]{Halzen:1984mc}
\bibinfo{author}{\bibfnamefont{F.}~\bibnamefont{Halzen}} \bibnamefont{and}
  \bibinfo{author}{\bibfnamefont{A.~D.} \bibnamefont{Martin}},
  \emph{\bibinfo{title}{{Quarks and leptons: An introductory course in modern
  particle physics}}} (\bibinfo{publisher}{Wiley}, \bibinfo{address}{New York},
  \bibinfo{year}{1984}).

\bibitem[{\citenamefont{Berestetskii et~al.}(1982)\citenamefont{Berestetskii,
  Lifshitz, and Pitaevskii}}]{Berestetsky:1982aq}
\bibinfo{author}{\bibfnamefont{V.~B.} \bibnamefont{Berestetskii}},
  \bibinfo{author}{\bibfnamefont{E.~M.} \bibnamefont{Lifshitz}},
  \bibnamefont{and} \bibinfo{author}{\bibfnamefont{L.~P.}
  \bibnamefont{Pitaevskii}}, \emph{\bibinfo{title}{{Quantum Electrodynamics}}},
  vol.~\bibinfo{volume}{4} of \emph{\bibinfo{series}{Course of Theoretical
  Physics}} (\bibinfo{publisher}{Butterworth-Heinemann},
  \bibinfo{address}{Oxford}, \bibinfo{year}{1982}).

\bibitem[{\citenamefont{Sjostrand et~al.}(2006)\citenamefont{Sjostrand, Mrenna,
  and Skands}}]{Sjostrand:2006za}
\bibinfo{author}{\bibfnamefont{T.}~\bibnamefont{Sjostrand}},
  \bibinfo{author}{\bibfnamefont{S.}~\bibnamefont{Mrenna}}, \bibnamefont{and}
  \bibinfo{author}{\bibfnamefont{P.~Z.} \bibnamefont{Skands}},
  \bibinfo{journal}{JHEP} \textbf{\bibinfo{volume}{05}}, \bibinfo{pages}{026}
  (\bibinfo{year}{2006}).

\bibitem[{\citenamefont{Lange}(2001)}]{Lange:2001uf}
\bibinfo{author}{\bibfnamefont{D.~J.} \bibnamefont{Lange}},
  \bibinfo{journal}{Nucl. Instrum. Meth.} \textbf{\bibinfo{volume}{A462}},
  \bibinfo{pages}{152} (\bibinfo{year}{2001}).

\bibitem[{\citenamefont{Biryukov et~al.}(1997)}]{Biryukov1997}
\bibinfo{author}{\bibfnamefont{V.~M.} \bibnamefont{Biryukov}}
  \bibnamefont{et~al.}, \emph{\bibinfo{title}{{Crystal Channeling and Its
  Application at High-Energy Accelerators}}}
  (\bibinfo{publisher}{Springer-Verlag}, \bibinfo{address}{Berlin},
  \bibinfo{year}{1997}).

\bibitem[{sup()}]{supplemental}
\bibinfo{howpublished}{See Supplemental Material, which includes Refs.~\cite{Thomas:1926dy,Thomas:1927yu,Bargmann:1959gz}.}

\bibitem[{\citenamefont{Kendall et~al.}(1983)\citenamefont{Kendall, Stuart, and
  Ord}}]{Kendall:1983}
\bibinfo{author}{\bibfnamefont{M.}~\bibnamefont{Kendall}},
  \bibinfo{author}{\bibfnamefont{A.}~\bibnamefont{Stuart}}, \bibnamefont{and}
  \bibinfo{author}{\bibfnamefont{J.}~\bibnamefont{Ord}},
  \emph{\bibinfo{title}{The advanced theory of statistics}}
  (\bibinfo{publisher}{Charles Griffin}, \bibinfo{address}{London},
  \bibinfo{year}{1983}).

\bibitem[{\citenamefont{Davier et~al.}(1993)\citenamefont{Davier, Duflot,
  Le~Diberder, and Rouge}}]{Davier:1992nw}
\bibinfo{author}{\bibfnamefont{M.}~\bibnamefont{Davier}},
  \bibinfo{author}{\bibfnamefont{L.}~\bibnamefont{Duflot}},
  \bibinfo{author}{\bibfnamefont{F.}~\bibnamefont{Le~Diberder}},
  \bibnamefont{and} \bibinfo{author}{\bibfnamefont{A.}~\bibnamefont{Rouge}},
  \bibinfo{journal}{Phys. Lett.} \textbf{\bibinfo{volume}{B306}},
  \bibinfo{pages}{411} (\bibinfo{year}{1993}).

\bibitem[{\citenamefont{Voss et~al.}(2007)\citenamefont{Voss, Hoecker, Stelzer,
  and Tegenfeldt}}]{Hocker:2007ht}
\bibinfo{author}{\bibfnamefont{H.}~\bibnamefont{Voss}},
  \bibinfo{author}{\bibfnamefont{A.}~\bibnamefont{Hoecker}},
  \bibinfo{author}{\bibfnamefont{J.}~\bibnamefont{Stelzer}}, \bibnamefont{and}
  \bibinfo{author}{\bibfnamefont{F.}~\bibnamefont{Tegenfeldt}},
  \bibinfo{journal}{PoS} \textbf{\bibinfo{volume}{ACAT}}, \bibinfo{pages}{040}
  (\bibinfo{year}{2007}).

\bibitem[{\citenamefont{Mazzolari}(2018)}]{Mazzolari:2018}
\bibinfo{author}{\bibfnamefont{A.}~\bibnamefont{Mazzolari}},
  \bibinfo{howpublished}{(private communication)} (\bibinfo{year}{2018}).

\bibitem[{\citenamefont{Apollinari et~al.}(2017)\citenamefont{Apollinari,
  Bejar~Alonso, Bruning, Fessia, Lamont, Rossi, and
  Tavian}}]{ApollinariG.:2017ojx}
\bibinfo{author}{\bibfnamefont{G.}~\bibnamefont{Apollinari}},
  \bibinfo{author}{\bibfnamefont{I.}~\bibnamefont{Bejar~Alonso}},
  \bibinfo{author}{\bibfnamefont{O.}~\bibnamefont{Bruning}},
  \bibinfo{author}{\bibfnamefont{P.}~\bibnamefont{Fessia}},
  \bibinfo{author}{\bibfnamefont{M.}~\bibnamefont{Lamont}},
  \bibinfo{author}{\bibfnamefont{L.}~\bibnamefont{Rossi}}, \bibnamefont{and}
  \bibinfo{author}{\bibfnamefont{L.}~\bibnamefont{Tavian}},
  \emph{\bibinfo{title}{{\CHc High-Luminosity Large Hadron Collider (HL-LHC)}}}
  (\bibinfo{publisher}{CERN Yellow Rep. Monogr. {\bf 4}},
  \bibinfo{year}{2017}).

\bibitem[{\citenamefont{Thomas}(1926)}]{Thomas:1926dy}
\bibinfo{author}{\bibfnamefont{L.~H.} \bibnamefont{Thomas}},
  \bibinfo{journal}{Nature} \textbf{\bibinfo{volume}{117}},
  \bibinfo{pages}{514} (\bibinfo{year}{1926}).

\bibitem[{\citenamefont{Thomas}(1927)}]{Thomas:1927yu}
\bibinfo{author}{\bibfnamefont{L.~H.} \bibnamefont{Thomas}},
  \bibinfo{journal}{Phil. Mag.} \textbf{\bibinfo{volume}{3}},
  \bibinfo{pages}{1} (\bibinfo{year}{1927}).

\bibitem[{\citenamefont{Bargmann et~al.}(1959)\citenamefont{Bargmann, Michel,
  and Telegdi}}]{Bargmann:1959gz}
\bibinfo{author}{\bibfnamefont{V.}~\bibnamefont{Bargmann}},
  \bibinfo{author}{\bibfnamefont{L.}~\bibnamefont{Michel}}, \bibnamefont{and}
  \bibinfo{author}{\bibfnamefont{V.~L.} \bibnamefont{Telegdi}},
  \bibinfo{journal}{Phys. Rev. Lett.} \textbf{\bibinfo{volume}{2}},
  \bibinfo{pages}{435} (\bibinfo{year}{1959}).

\end{thebibliography}

\onecolumngrid
\newpage

\begin{center}
{\large \bfseries \boldmath
Novel method for the direct measurement of the \Ptau lepton dipole moments}\\
\end{center}

\begin{center}
The following includes Supplemental Material for the electronic version.
\end{center}

\vskip1cm

The time evolution of the spin-polarization vector {\bf s} is regulated by the T-BMT equation~\cite{Thomas:1926dy,Thomas:1927yu,Bargmann:1959gz}.
The precession of the spin-polarization vector induced by the interaction of the MDM and the EDM of a
charged particle channeled in a bent crystal is derived in Refs.~\cite{Botella:2016ksl,Bagli:2017foe},
assuming $s_{0,\yaxis}$ initial polarization and $(g-2) \gg 1/\gamma,~\dt$. 
For the \Ptau lepton $\at = (g-2)/2 \approx 10^{-3} \sim 1/\gamma$, and initial $s_{0,\zaxis}$ and $s_{0,\yaxis}$ polarizations are possible.
Under these conditions the spin equation of motion reads
\begin{align} 
s_\xaxis & \approx s_{0,\zaxis} \frac{\dpt}{\apdt}\left[ - c \sin\Phi + \frac{\apt}{\apdt}s\left(1-\cos\Phi\right) \right]
                  + s_{0,\yaxis} \frac{\dpt}{\apdt}\left[ \frac{\apt}{\apdt} c \left( 1 - \cos\Phi \right) + s\sin\Phi \right] , \nonumber \\   
%          \approx - s_{0,\zaxis} \frac{\dpt}{\apdt}\sin\Phi
%                  + s_{0,\yaxis} \frac{\dpt\apt}{\apdt^2} \left( 1-\cos\Phi \right) , \nonumber \\  
s_\yaxis & \approx   s_{0,\zaxis} \frac{\apt}{\apdt}\left[ \sin\Phi + \frac{\dpt^2}{\apdt\apt}sc\left(1-\cos\Phi\right) \right]
                  + s_{0,\yaxis} \frac{\apt^2}{\apdt^2}\left[ \frac{\dpt^2}{\apt^2}c^2 + \left( 1 + \frac{\dpt^2}{\apt^2}s^2 \right)\cos\Phi  \right] , \nonumber \\
%          \approx s_{0,\zaxis} \frac{\apt}{\apdt}\sin\Phi
%                  s_{0,\yaxis} \frac{\dpt^2}{\apdt^2} + \frac{\apt^2}{\apdt^2} \cos\Phi , \nonumber \\
s_\zaxis & \approx  s_{0,\zaxis} \frac{\apt^2}{\apdt^2}\left[ \left( 1 + \frac{\dpt^2}{\apt^2}c^2 \right)\cos\Phi + \frac{\dpt^2}{\apt^2}s^2 \right]
                 + s_{0,\yaxis} \frac{\apt}{\apdt}\left[ - \sin\Phi + \frac{\dpt^2}{\apdt\apt}sc \left( 1 - \cos\Phi \right) \right] ,
%          \approx  s_{0,\zaxis} \cos\Phi
%                 - s_{0,\yaxis} \frac{\apt}{\apdt} \sin\Phi ,
\label{eq:eom_main_szsy_precise}
\end{align}
where
$\apt = \at+\frac{1}{1+\gamma}$,
$\dpt = \dt/2$,
$\apdt = \sqrt{\apt^2+\dpt^2}$,
and $\Phi = \gamma \theta_C \apdt$ is the precession angle, with $\theta_C$ the crystal bending angle.
The coefficients $s$ and $c$ are given by $\sin(\overline{\Omega t})$ and $\cos(\overline{\Omega t})$, respectively,
where $\overline{\Omega t} \approx c/\rho_0 \times L/c = \theta_C \sim 10^{-2}$, with $\rho_0$ ($L$) the crystal curvature radius (length),
is the average rotation angle of the particle trajectory when traversing the bent crystal with revolution frequency $\Omega$ in a time interval $t$.
These expressions hold at precision ${\cal O}(10^{-5})$. Approximating $s\approx 0$ and $c\approx 1$ we obtain
Eq.~(4) reported in the Letter,
which applies at ${\cal O}(10^{-2})$.
In the limit $(g-2) \gg 1/\gamma, \dt$ and $s_{0,\zaxis} = 0$,
it
reduces to
\begin{align}
s_\xaxis & 
                     \approx s_{0,\yaxis} \frac{\dt}{g-2} \left( 1-\cos\Phi \right) , \nonumber \\  
s_\yaxis & 
                     \approx s_{0,\yaxis} \cos\Phi, \nonumber \\
s_\zaxis & 
                     \approx - s_{0,\yaxis} \sin\Phi ,
\label{eq:eom_sybaryons}
\end{align}
with $\Phi \approx \frac{g-2}{2}\gamma\theta_C$.

\begin{figure}[htb]
\centering
\begin{tabular}{cc}
\includegraphics[width=0.35\textwidth]{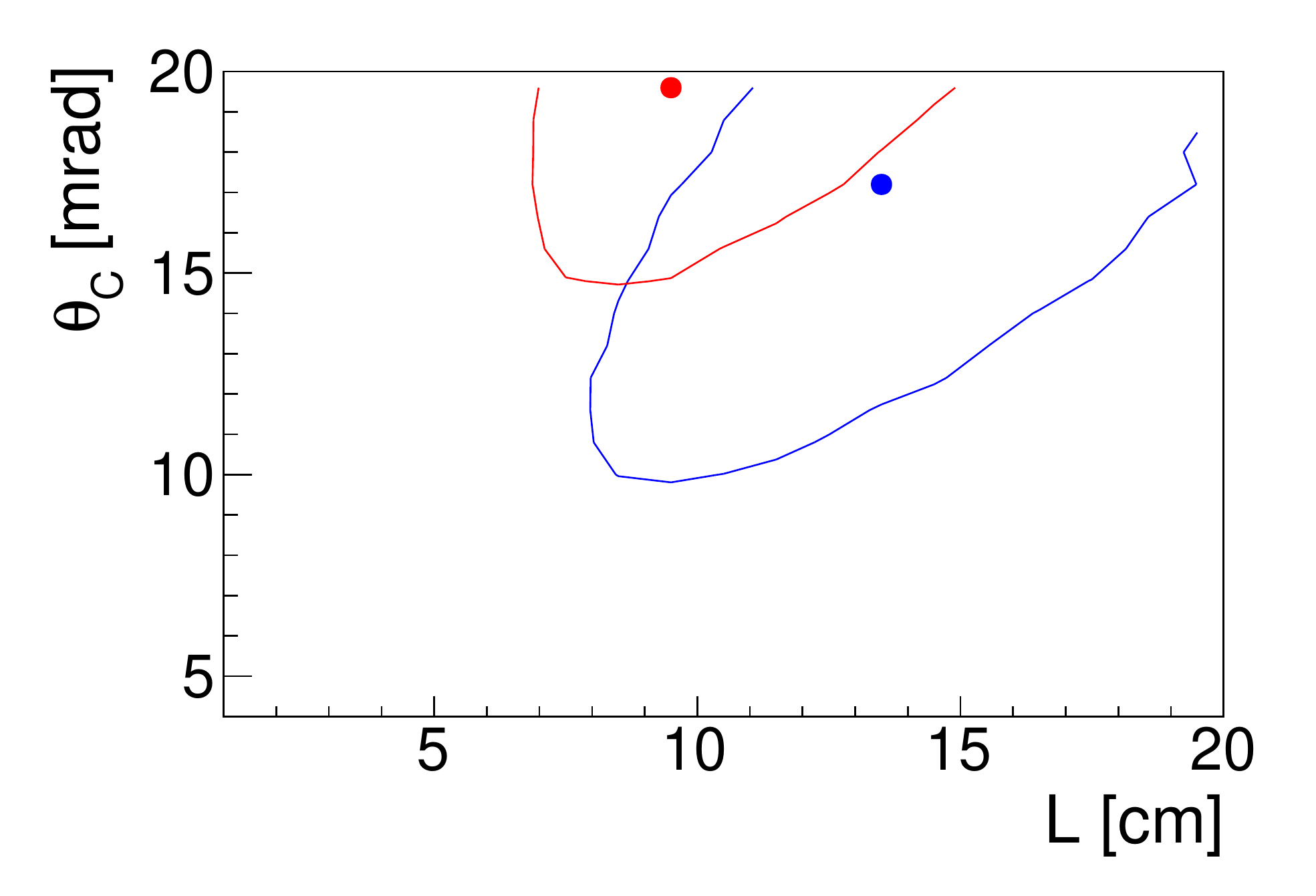}
\includegraphics[width=0.35\textwidth]{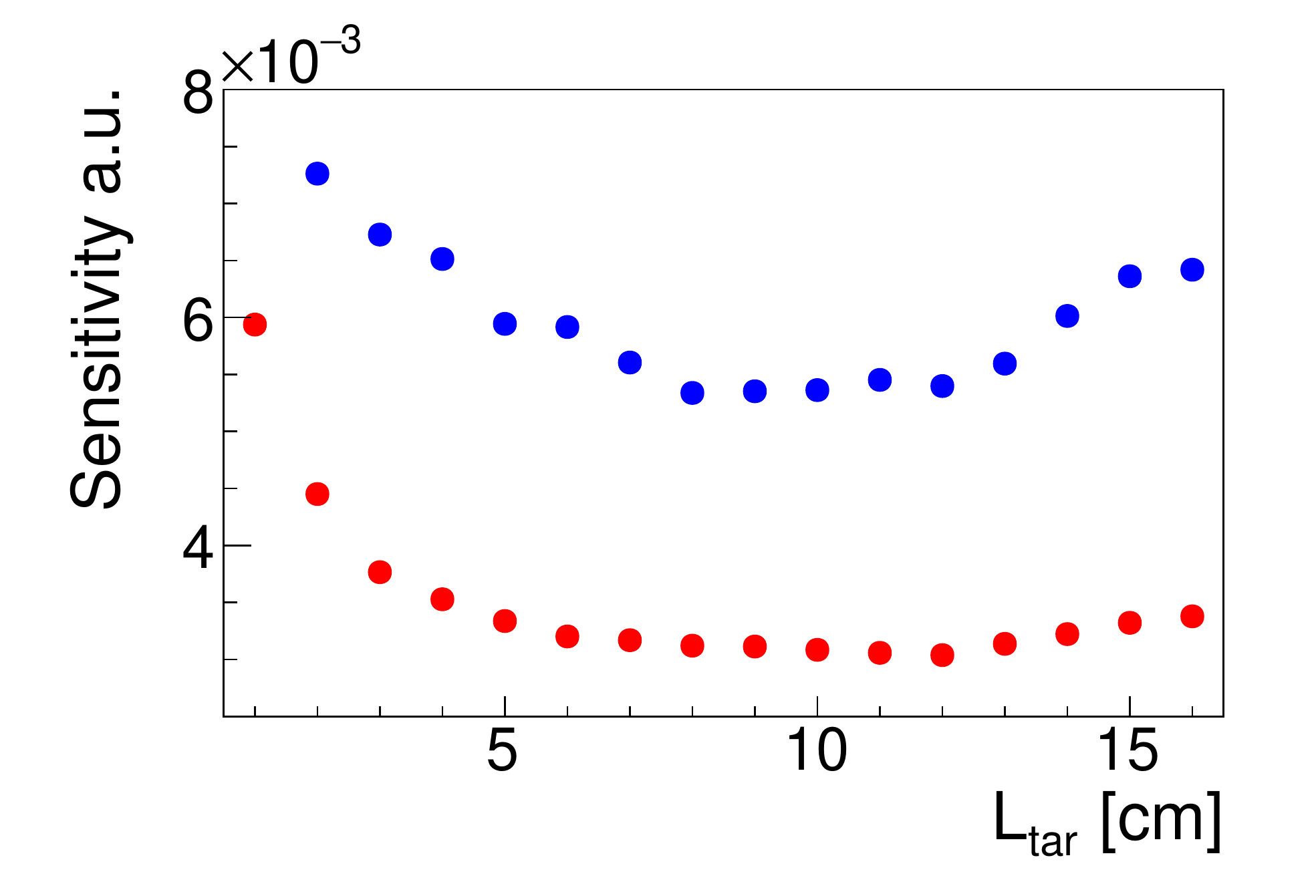}
\end{tabular}
\caption{\label{fig:optimization_supp} %(color online).
  Regions of minimal uncertainty for both \at and \dt as a function of the crystal parameters \Lc and \thc (left) and \Ltarc (right), for \Ge (red) and \Si (blue),
  for initial  $s_{0,\zaxis}$ polarization. In the left figure the lines represent regions whose uncertainties on \at and \dt are increased by 10\% with respect to the minimum (points).
}
\end{figure}

\begin{figure}[htb]
\centering
\begin{tabular}{cc}
\includegraphics[width=0.35\textwidth]{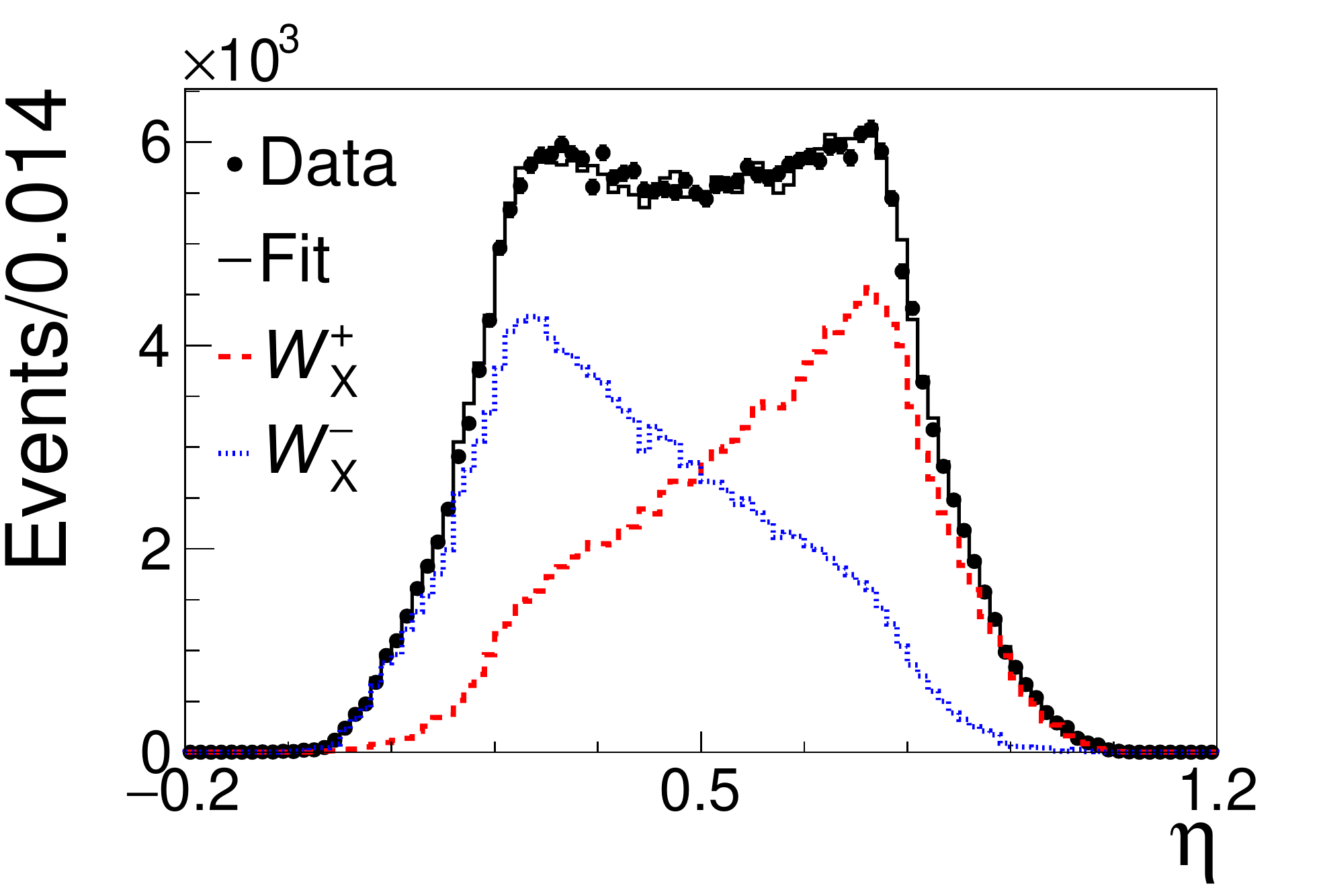}
\includegraphics[width=0.35\textwidth]{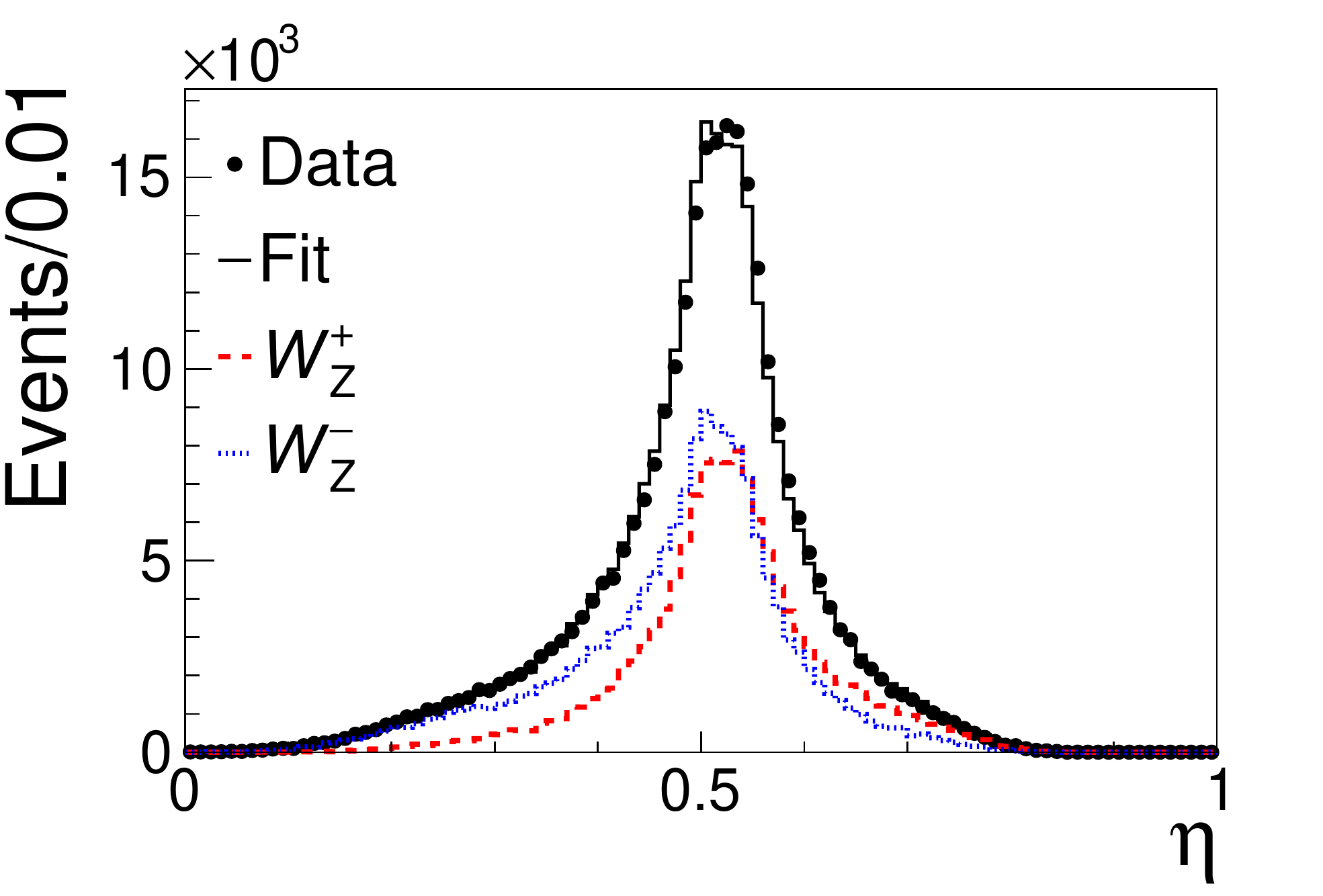}
\end{tabular}
\caption{\label{fig:template_fit_supp} %(color online).
Template fit results to the classifier response using simulated events for $s_{\xaxis}$ (left)
 and $s_{\zaxis}$ (right) polarizations.
 The normalized separation between \Wipeta (dashed red lines) and \Wimeta (dotted blue lines) determines the corresponding average event information $S_\xaxis\approx 0.42$ (left) and $S_\zaxis\approx 0.23$ (right).
}
\end{figure}

\end{document}